\documentclass[twocolumn]{aa}
\usepackage{graphicx}
\begin{document}
   \title{Photometric study of the young open cluster NGC 3293
          \thanks{Based on observations collected at UTSO, ESO (Dutch 0.9 m 
	          telescope) and CASLEO. The CCD and data acquisition system at
                  CASLEO has been partly financed by R.M. Rich through U.S. NSF 
                  Grant AST-90-15827}
         }
   \subtitle{}

   \author{{}G. Baume\inst{1,2}
             \fnmsep\thanks{Visiting Astronomer, CASLEO operated under
	     agreement between CONICET of the Rep\'ublica Argentina and the
	     National Universities of La Plata, C\'ordoba and San Juan},
             R. A. V\'azquez\inst{1,**},
             G. Carraro\inst{2}
             \and 
	     A. Feinstein\inst{1}
          }

   \offprints{gbaume@fcaglp.fcaglp.unlp.edu.ar}

   \institute{Facultad de Ciencias Astron\'{o}micas y Geof\'{\i}sicas de la
              UNLP, IALP-CONICET, Paseo del Bosque s/n, 1900, La Plata, 
	      Argentina\\
         \and
             Dipartimento di Astronomia, Universit\`{a} di Padova,
             Vicolo Osservatorio 2, I-35122 Padova, Italy\\
             }

\date{Received **; accepted ** }

\abstract{{}
Deep and extensive CCD photometric observations $UBV(RI)_{C}H_{\alpha}$ were 
carried out in the area of the open cluster NGC 3293. The new data set allows
to see the entire cluster sequence down to $M_{V} \approx +4.5$, revealing that 
stars with $M_{V} < -2$ are evolving off the main sequence; stars with $-2 < 
M_{V} < +2$ are located on the main sequence and stars with $M_{V} > +2$ are 
placed above it. According to our analysis, the cluster distance is $d = 
2750 \pm 250~pc$ ($V_{0}-M_{V} = 12.2 \pm 0.2$) and its nuclear age is 
$8 \pm 1~Myr$. NGC 3293 contains an important fraction of pre--main sequence 
(PMS) stars distributed along a parallel band to the ZAMS with masses from $1$ 
to $2.5 \cal M_{\sun}$ and a mean contraction age of $10~Myr$. This last value 
does not differ too much from the nuclear age estimate. Actually, if we take 
into account the many factors that may affect the PMS star positions onto the 
colour--magnitude diagram, both ages can be perfectly reconciled. The star 
formation rate, on the other hand, suggests that NGC 3293 stars formed surely 
in one single event, therefore favouring a coeval process of star formation. 
Besides, using the $H_{\alpha}$ data, we detected nineteen stars with signs of 
having $H_{\alpha}$ emission in the region of NGC 3293, giving another 
indication that the star formation process is still active in the region. The 
computed initial mass function for the cluster has a slope value $x = 1.2 \pm 
0.2$, a bit flatter than the typical slope for field stars and similar to the 
values found for other young open clusters.

}

\authorrunning{Baume et al.}
\titlerunning{NGC 3293}

\maketitle

\section{Introduction}

$\hspace{0.5cm}$
Star clusters constitute the most appropriate laboratory to test the stellar 
evolution theory since all the stars formed in a cluster belong to the same 
region of space, are all at the same distance and have the same chemical 
composition. Some of the main tools, tightly related to the history of the star 
formation processes, are the luminosity function (LF) and the initial mass 
function (IMF). Regarding the construction of these distributions, open clusters 
offer two operational advantages when compared to field stars: a) there is no 
need to assume a time-independent IMF as necessary in deriving the IMF field 
stars and, b) as cluster stars do not move far from their birth--sites, neither 
is necessary to consider the IMF is spatially independent (Scalo 1986; Herbst \& 
Miller 1982; Wilner \& Lada 1991). Still, after several years of intensive and 
extensive works on open clusters (in our galaxy and the Magellanic Clouds) 
applying the powerful CCD techniques, some questions have not found the answer 
yet:
\begin{itemize}
\item  Which is the star formation mechanism and how long it keeps 
       active (Sung et al. 1998)? Is it coeval or sequential (Iben \& Talbot 
       1966; Herbst \& Miller 1982; Adams et al. 1983)?
\item  Is the IMF slope universal? Is it bimodal? Why do some open clusters 
       show an apparent deficiency of low mass stars (van den Bergh \& Sher 
       1960; Adams et al. 1983; Lada et al. 1993; Phelps \& Janes 1993; Sung et 
       al. 1998; Prisinzano et al. 2001), despite they are young enough to
       exclude stellar losses by dynamical evolution? 
\item  Does the IMF vary from cluster to cluster even within a same star 
       formation region probably by changes in the initial conditions of the 
       star formation process (Scalo 1986; Lada \& Lada 1995)?
\end{itemize}

The young open cluster NGC 3293 = C1033--579 ($l = 285.9^{\circ}$, 
$b = 0.07^{\circ}$) is placed in the Carina region north--west of Trumpler 
14/16. The three clusters are embedded in the nebulosity of the HII region NGC 
3372, although NGC 3293 is relatively free of patches of dust. This cluster is 
compact, well populated and not too much reddened. Besides, with an age 
$< 10~Myr$, it is surely free from dynamical evolution too. Altogether these 
properties make it an excellent target to examine some of the items enumerated 
above. Herbst \& Miller (1982, hereafter HM82) performed the deepest photometric 
study (mostly photographic) of this object to investigate the star--forming 
history down to $V \approx 15$. HM82 found features that deserve confirmation: 
a) the low mass stars formed first, b) the cluster LF shows a sharp dip at 
$M_{V} = +2$ and $+3$, c) the cluster IMF is not only flatter than the field 
star IMF but it varies with time during the cluster formation period and d) the 
cluster has a halo structure formed by less massive stars. The need for new deep 
photometry is evident as a substantial part of the HM82's findings may have been 
produced by selection effects close to the detection limit of the photographic 
plates (Deeg \& Ninkov 1996). \\

Apart from the HM82 study of the star--forming history, a synthesis of the
main investigations carried out in NGC 3293 includes: spectroscopy of bright 
cluster members to obtain radial velocities and spectral classification by Feast 
(1958); rotational velocity studies undertaken by Balona (1975); $UBVRI$ 
photoelectric photometry and polarimetric measures performed by Feinstein \& 
Marraco (1980, hereafter FM80); $UBV$ photometry including the cluster and the 
surrounding area was also carried out by Turner et al. (1980, hereafter TGHH80); 
important contributions including the detection of several $\beta$ Cepheid stars 
come from $ubvyH_{\beta}$ observations (Shobbrook, 1980; Balona \& Engelbrecht 
1981, 1983; Shobbrook 1983; Balona 1994). \\

\begin{figure*}
  \centering
   \includegraphics[width=14cm]{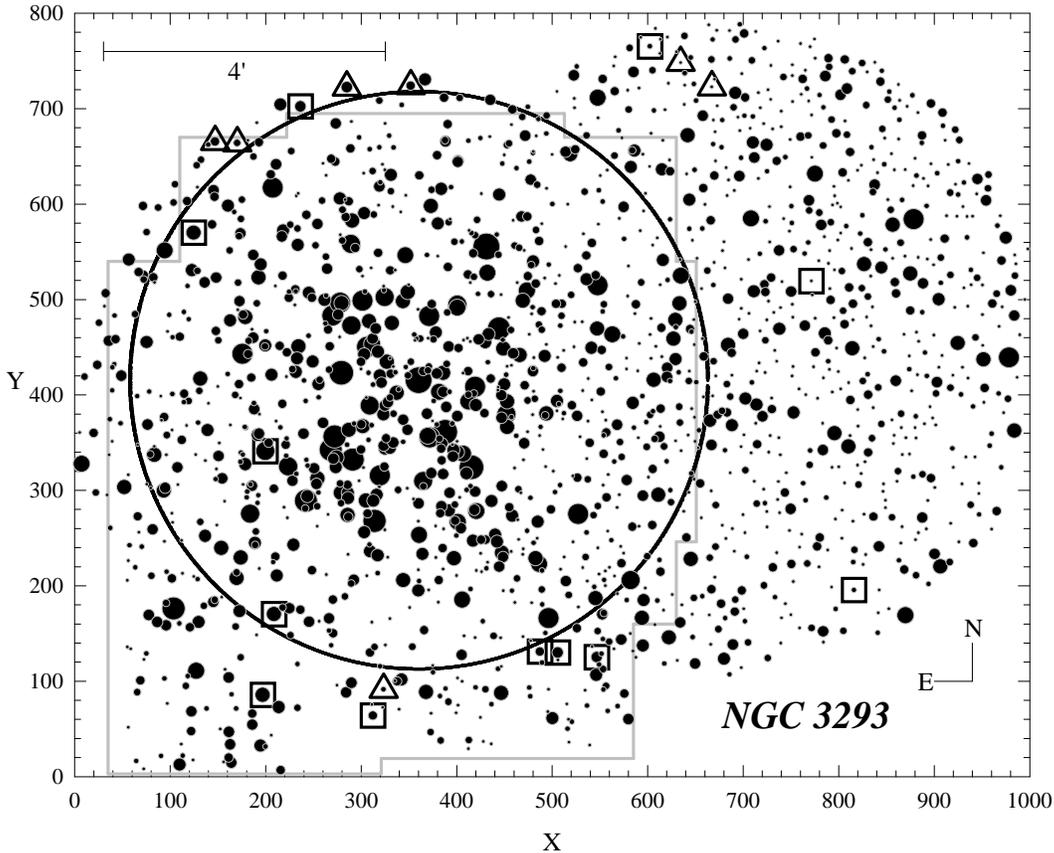}
      \caption{Finding chart of the observed area ($V$ filter). Black circle,
      $4'\!.1$ radius, indicates the adopted limits for the cluster from section
      3.1. Grey lines indicates limits of frames from UTSO and ESO observations
      (that is stars with $U$ and $B$ measurements). For a coordinate reference,
      star \#2 ($X = 360.3$; $Y = 415.4$), adopted as the cluster centre, has
      $\alpha_{2000} = 10^{h}~35^{m}~49^{s}\!.3$; $\delta_{2000} = 
      -58^{\circ}~13^{'}~27^{''}\!\!.4$. Stars enclosed by squares and triangles 
      are likely and probable $H_{\alpha}$ emission stars respectively (see 
      section 5).}
      \label{Fig01}
\end{figure*}

The present investigation aims at defining the lower main sequence structure of 
this cluster and detecting the presence of pre--main sequence (PMS) stars. We 
understand that a primary (but not concluding) indicator of the existence of PMS 
objects comes from the detection of faint cluster members above the ZAMS (Walker 
1957, 1961). More recent evidences on PMS stars (cf. Fig. 3 in Preibisch \& 
Zinnecker 1999) indicate that they conform a band $1^{m}$ above the ZAMS, $1$ to 
$2^{m}$ wide approximately in the HR diagram. However, the determination of this 
locus through photometry alone is completely spoiled because of the 
contamination by field interlopers. Better defined locations are achieved by 
removing the field star contamination through adequate comparison fields. That 
implies to obtain a ``clean'' colour--magnitude diagram (CMD), where the PMS 
stars, if they do exist, occupy a well defined place. Another purpose of this 
investigation deals with the analysis of the frequency distribution of both, 
magnitudes and masses, including not only the brightest (most massive) members 
but also the faintest ones. Finally we will attempt to detect $H_{\alpha}$ 
emission stars (indicative of the PMS stage) through an $H_{\alpha}(on/off)$ 
survey that was conducted in the NGC 3293 region. \\

In section 2 we describe our observations, the reduction procedure and the 
complementary sources of information. Section 3 contains the data analysis 
including the estimation of the cluster angular size and the membership 
assignment. Section 4 illustrates the determination of the basic cluster 
parameters: reddening, distance, linear size and age; it also contains the 
cluster LF and IMF determination. Section 5 describes the process to detect 
stars with $H_{\alpha}$ emission. In section 6 we discuss the star formation 
process. Finally, the conclusions are included in section 7 along with a 
summary of our main findings. \\

\section{Data Set} 

$\hspace{0.5cm}$
The main data set comes from CCD photometric observations of stars in the region 
of NGC 3293 carried out along several observational runs complemented with 
available information from the literature. Data come then from:
\begin{itemize}
\item[1.] The University of Toronto Southern Observatory (UTSO): In two 
          observational runs we obtained $UBV(RI)_{C}$ photometry using a 
	  PM--512 METHACROME UV coated CCD (scale $0''\!\!.45/pixel$ covering 
	  $4'$ on a side) attached to the Hellen--Sawyer 60--cm telescope. The 
	  first run was in 1994 April 13, 14 and 16 when three frames  were 
	  exposed in NGC 3293 using the  nitrogen--cooled detector; the second 
	  run took place in 1996 February 25 and 26, when three more frames were 
	  exposed using a glycol--refrigerated detector. Short (2 to 6 s), mid 
	  (100 to 200 s) and long (two series of up to 1100 s) exposure times 
	  were used to get photometry of the bright stars and to improve the 
	  signal--noise--ratio of the faintest stars respectively. A $BV$ 
	  comparison frame was taken on February 27, $20'$ north of the cluster, 
	  using similar exposure times. Weather conditions at UTSO were always 
	  photometric with seeing values ranging from $1''\!\!.1$ to 
	  $1''\!\!.5$. 
\item[2.] The Complejo Astron\'{o}mico El Leoncito (CASLEO): In 1999 April 15 and 17 
          we made $V(R)_{C}H_{\alpha}$ observations in two frames ($4'\!.5$ 
	  radius) in the area of NGC 3293 using a nitrogen--cooled detector 
	  Tek--1024 CCD and focal reducer attached to the 215--cm telescope 
	  (scale $0''\!\!.813/pixel$). One frame was centred in the cluster and 
	  the other west side of it (see Fig. 1). Exposure times in $V$, 
	  $H_{\alpha}(on) = 656.6~nm$ and $H_{\alpha}(off) = 666.0~nm$ filters 
	  ranged from $1~s$ (in central frame) or $15~s$ (west side frame) to 
	  $700~s$, and from $1$ to $150~s$ for the $R$ band. In all the cases, 
	  mid exposure frames were also taken. 
\item[3.] The European Southern Observatory (ESO): We complemented our data set  
          with an unpublished $UBV(RI)_{C}$ CCD photometric survey conducted by 
	  F. Patat and G. Carraro in 1996 at ESO 0.9 m Dutch telescope. Details 
	  of these observations and data reductions are given in Patat \& 
	  Carraro (2001). 
\item[4.] Other data sources: photometric data for a few bright stars (see Table 
          1) and available spectral classification were taken from FM80 and 
	  TGHH80. Useful complementary information was mainly derived from the 
	  Tycho Catalog (ESA 1997) and $SIMBAD$ databases (see Tables 1 and 3).
\end{itemize}

Figure 1 shows the finding chart of all measured stars. The grey line stands for 
the area surveyed at UTSO and ESO ($UBV(RI)_{C}$) and the black circle encloses 
the $''cluster~region''$ (see section 3.1). \\

   \begin{table*}
      \caption[]{Photometric catalogue of the NGC 3293 region.}
      \fontsize{8} {10pt}\selectfont
      \begin{tabular}{cccccrrr@{.}lr@{.}lr@{.}lr@{.}lr@{.}lr@{.}ll@{ - }l}
      \hline
        $\#$ & $\#_{T}$ & $\#_{FM}$ & $\#_{F}$ & $\#_{HM}$ & $X~~~$ & $Y~~~$ & \multicolumn{2}{c}{$V~~$} & \multicolumn{2}{c}{$U-B~$} & \multicolumn{2}{c}{$B-V~~~$} & \multicolumn{2}{c}{$V-R~~~$} & \multicolumn{2}{c}{$V-I~~~$} & \multicolumn{2}{c}{{}$\Delta~H_{\alpha}$} & \multicolumn{2}{c}{Remarks}  \\
      \hline
    1 &   3 &   3 &   3 &  -  &  431.2 &  555.4 &  6&52                 & -0&81                 &  0&12                  &  0&07 $_{FM}$         &  0&09 $_{FM}$         &   &                   & lm1 & HD 91943        \\
    2 &   4 &   4 &   4 &  -  &  360.3 &  415.4 &  6&54                 & -0&82                 &  0&00                  &  0&04 $_{FM}$         &  0&09 $_{FM}$         &   &                   & lm1 & HD 91969        \\
    : &   : &   : &   : &  :  &   :~~~ &   :~~~ & \multicolumn{2}{c}{:} & \multicolumn{2}{c}{:} &  \multicolumn{2}{c}{:} & \multicolumn{2}{c}{:} & \multicolumn{2}{c}{:} & \multicolumn{2}{c}{:} & \multicolumn{2}{c}{:} \\
 1689 &  -  &  -  &  -  &  -  &  339.9 &   70.4 & 21&65 ::              &   &                   &  1&12 ::               &   &                   &   &                   &   &                   & --  & --              \\
 1690 &  -  &  -  &  -  &  -  &  585.0 &  131.4 & 21&72 ::              &   &                   &  0&89 ::               &   &                   &   &                   &   &                   & --  & --              \\
      \hline
      \end{tabular}\\
      \begin{tabular}{c}
      \begin{minipage}{16cm}
        {\bf Comment:} Table 1 is available in full in an electronic version at
	the CDS. A brief summary, only indicating its structure, is shown here.
      \end{minipage}
      \end{tabular}
   \end{table*}

\renewcommand{\thefootnote}{\dag}
The reduction process was carried out using IRAF\footnote{IRAF is distributed by 
NOAO, which are operated by AURA, under cooperatative agreement with NSF} 
CCDRED, DAOPHOT and PHOTCAL packages. Instrumental signatures at UTSO and CASLEO 
frames were removed using bias and dome flat exposures. Dark currents were 
tested to recognise its significance in our observations but they were found 
negligible. Instrumental magnitudes $UBV(RI)_{C}H_{\alpha}$ were produced via 
the point spread function, PSF, (Stetson 1987). Calibration sequences in the 
open clusters Hogg 16 and NGC 5606 (V\'azquez et al. 1991, 1994) including 
several blue and red stars and standard star groups with blue and red stars from 
Landolt (1992) were used to produce final colours and magnitudes at UTSO and 
CASLEO respectively. The final errors of the respective calibration equations, 
adopted as external errors of our photometry, were $< 0.025$. Small mean 
differences $< 0.03$ were found between UTSO and ESO measures. However, the 
$V(R)_{C}$ measures made at CASLEO showed a shift relative to UTSO--ESO 
photometry that was corrected to bring them into the system defined at 
UTSO--ESO. The estimate of the internal errors was done comparing colour and 
magnitudes of the stars located in the overlapping zones of our frames. That 
yields typical differences $< 0.03$, up to $V \approx 17$. Table 1 contains the 
photometric output for 1690 stars; it also includes the star identification, 
coordinates, the cross--references with other authors and some astronomical 
catalogues, and the membership assignment. Summarising the information available 
after our survey we have: 1690 stars with $V$ magnitude, 560 stars with $U-B$ 
index, 940 stars with $B-V$, 1550 with $V-R$, 903 with $V-I$ and 861 with 
$H_{\alpha}(on)-H_{\alpha}(off)$ index. \\

In relation to our data completeness, we performed an analysis for data
from UTSO (Baume 1999) using IRAF tasks ADDSTAR, DAOFIND and ALLSTAR. Then we
compared those results with ESO and CASLEO data. That analysis yielded us to
the following completeness results: $100 \%$ down to $V = 16$, $98.7 \%$ down to
$V = 17$, $94.3 \%$ down to $V = 18$ and $59.2 \%$ down to $V = 19$. \\

A comparison of our photometry with TGHH80, FM80 and HM82, in the sense ``our
photometry minus theirs'' is shown in Table 2. The influence of 
(photoelectrically) unresolved stars, binary and variable stars is reflected in 
the large standard deviations of the mean differences. As shown in the second 
rows, if the known anomalous stars are discarded, the deviations decrease 
substantially and the mean differences and standard deviations reach acceptable 
values. The exception is for HM82 photometry where the bulk of their data is 
photographic. \\

   \begin{table*}
      \centering
      \caption[]{Differences with previous photometric works in  the sense ``our photometry minus other authors''}
      \fontsize{8} {10pt}\selectfont
      \begin{tabular}{l|r@{$~\pm~$}lr@{$~\pm~$}lr@{$~\pm~$}lr@{$~\pm~$}lr@{$~\pm~$}lc}
      \hline
         Work  & \multicolumn{2}{c}{$\Delta(V)$} & \multicolumn{2}{c}{$\Delta(U-B)$} & \multicolumn{2}{c}{$\Delta(B-V)$} & \multicolumn{2}{c}{$\Delta(V-R)$} & \multicolumn{2}{c}{$\Delta(V-I)$} & $N$ \\
      \hline
         FM80  & -0.06 & 0.21 & -0.01 & 0.19 &  0.02 & 0.04 & -0.03 & 0.05 & -0.04 & 0.07 &  (35) \\
               & -0.01 & 0.03 &  0.03 & 0.05 &  0.03 & 0.03 & -0.03 & 0.03 & -0.03 & 0.06 &       \\
      \hline
        TGHH80 & -0.01 & 0.10 &  0.01 & 0.11 & -0.00 & 0.09 & \multicolumn{4}{c}{}        &  (86) \\
               & -0.01 & 0.06 &  0.01 & 0.08 & -0.00 & 0.04 & \multicolumn{4}{c}{}        &       \\
      \hline
        HM82   &  0.04 & 0.16 &  0.01 & 0.16 &  0.00 & 0.12 & \multicolumn{4}{c}{}        & (278) \\
      \hline
      \end{tabular}
   \end{table*}

\section{Data Analysis}

\subsection{Cluster angular radius}

$\hspace{0.5cm}$
To get reliable information on the evolutionary status of an open cluster we 
have to precisely know its size. In the present case, to address this issue we 
performed stellar counts in a $30' \times 30'$ Digitized Sky Survey (DSS) image 
centred in NGC 3293. All stars detected above a given threshold were assigned 
$V_{DSS}$ magnitudes using DAOPHOT task. $V_{DSS}$ magnitudes for stars brighter 
than $V \approx 18$ were next transformed into our photometric system with an 
accuracy of the order of $\pm 0.5^{m}$ (a procedure already applied in NGC 6231, 
Baume et al. 1999). Secondly, assuming the cluster is spherical, the highest 
apparent star density was fitted with a bi--dimensional Gaussian function. The 
cluster centre, defined by the highest peak (Battinelli et al. 1991), was found 
close to star \#2 (No. 4 in FM80 notation) and the angular radius was set at 
$4'\!.1$, the distance at which the star density equalises the background level. 
A similar procedure but using only our CCD $V$ data, that include stars as faint 
as $V \approx 20$ (Fig. 1), was applied. We counted stars inside concentric 
annuli, 100 pixels width, centred in star \#2. As at 650 pixels of star \#2, the 
annuli are not complete (only small portions of them were observed) we had to 
extrapolate the counts within each annulus to complete them. The stellar density 
profile coincides with the background density level at $4'\!.1$ from the cluster 
centre yielding the same result found with the DSS plates. The black circle of 
$4'\!.1$ radius shown in Fig. 1 defines then the area occupied by NGC 3293 
($''cluster~region''$). This agrees with TGHH80's finding who yielded a cluster 
angular diameter of $10'$. \\

\subsection{Cluster membership}

$\hspace{0.5cm}$
Proper motions and radial velocities studies yield the more accurate membership 
determination in a cluster. Up to now, proper motions are only feasible for 
nearby clusters (Sanner \& Geffert 2001) and radial velocities are available 
mostly for brightest stars. In NGC 3293, forty one of its brightest stars are 
listed in the Hipparcos/Tycho catalogues (ESA 1997). Only nine of them have 
parallax and proper motions with relative scientific value. They are indicated 
in Table 3 although the distance at which NGC 3293 is located makes them, 
statistically speaking, meaningless to derive distance and memberships.
Regarding radial velocities, we used Feast (1958) studies for membership
assignment to brightest stars. In particular, star \#3 (No. 21 in FM80 
notation) is a red super--giant ($CPD -57^{\circ}~3502$; IRAS source 10338-5759) 
and is a likely cluster member according to its radial velocity value and its 
spectrophotometric distance modulus (see Table 3). \\

In the classical photometric method to address cluster memberships (e.g. Baume 
et al. 1999), the consistency of the location of each star is assessed 
simultaneously in all the photometric diagrams (the two colour diagram, TCD, and 
the different CMDs). Other authors use reddening limits within which cluster
members should lie or they adopt maximum departures from a reference line as the 
ZAMS (Deeg \& Ninkov 1996; Hillenbrand et al. 1993). Our method works well for 
bright members, but it becomes unpractical to detect members among faint stars 
in crowded fields. It was mentioned by Abt (1979) that the classical method is 
controversial; however, as was stated by TGHH80 and emphasised by Forbes (1996), 
it is good enough when it relies on a careful inspection of the TCD and 
consistent reddening solutions are applied. The method works well too for 
nearby, not much reddened, intermediate and old clusters without traces of 
contracting stars (Stahler \& Fletcher 1991), but it fails for young clusters 
where contracting stars and field interlopers very often occupy the same 
location on the CMD. If the reddening is high and the cluster is distant, 
results are clearly dubious. Therefore, in order to minimise this problem, we 
divided our data set in two groups, stars brighter than $V \approx 14$ and 
fainter than that value. \\

\subsubsection{Bright cluster members}

\begin{figure*}
   \centering
   \includegraphics[width=7cm]{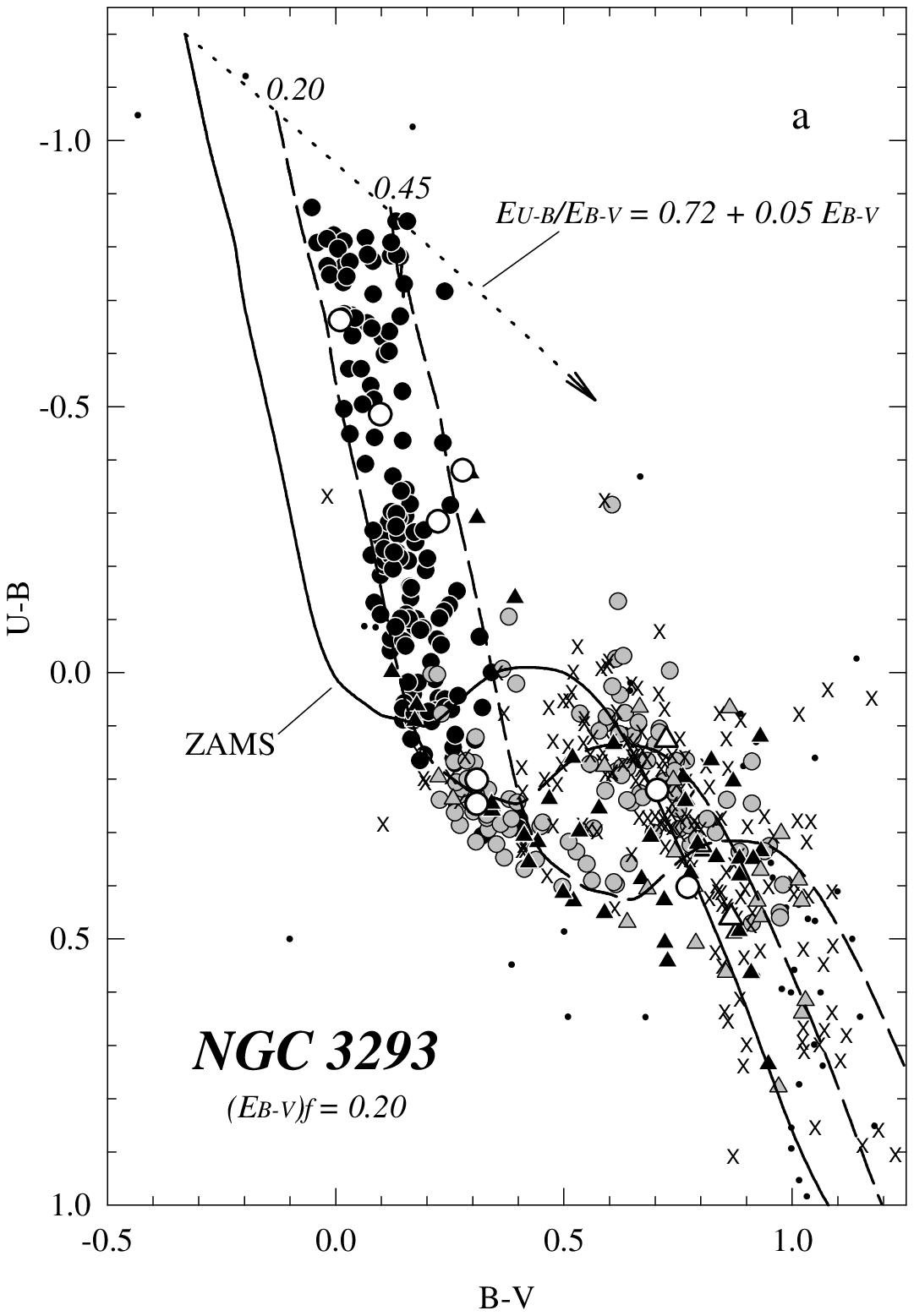}%
   \hspace{0.5cm}%
   \includegraphics[width=7cm]{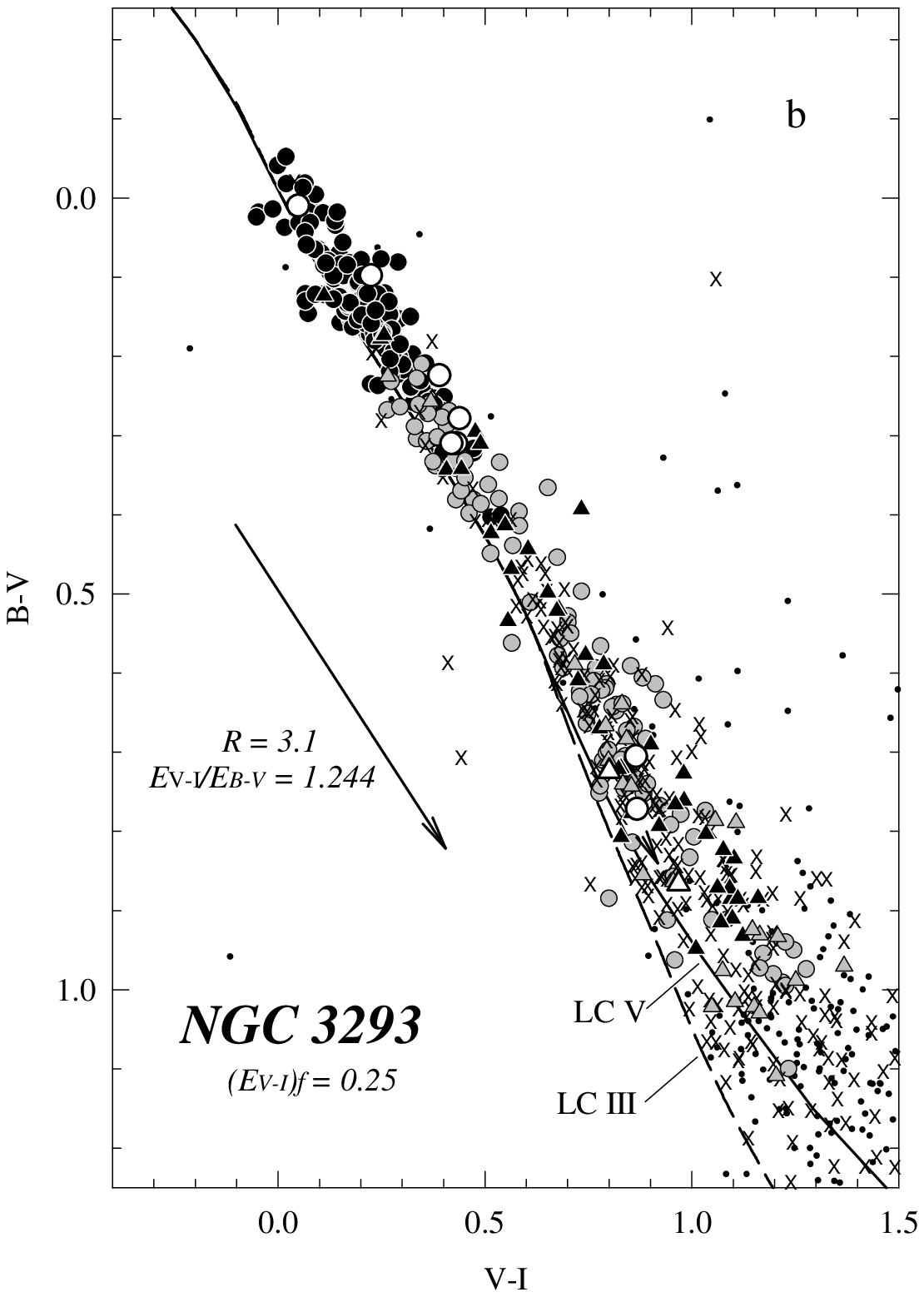}
      \caption{{} {\bf a)} Two colour diagram (TCD). Symbols have the following
      meaning: circles are likely member stars (lm) and triangles are probable
      member stars (pm); black symbols indicate membership obtained from the
      classical photometric method and grey ones those obtained from the
      subtraction method (see section 3.2 for explanation of the methods);
      crosses are non member stars (nm) and dots are stars without membership
      assignment; white symbols indicate likely and probable $H_{\alpha}$
      emission stars (circles and triangles respectively). Solid line is the
      Schmidt--Kaler's (1982) ZAMS; dashed lines indicate the position of the
      ZAMS shifted by $E_{B-V} = 0.20$ and $0.45$. Dotted arrow indicates the
      normal reddening path. {\bf b)} $B-V$ vs. $V-I$ diagram. Symbols as in
      Fig. 2a. Solid and dashed lines are the intrinsic positions for stars of
      luminosity classes V and III respectively (Cousins, 1978). Solid arrow
      gives the normal reddening path for $R = 3.1$}
      \label{Fig02}
\end{figure*}

\begin{figure*}
   \centering
   \includegraphics[width=14cm]{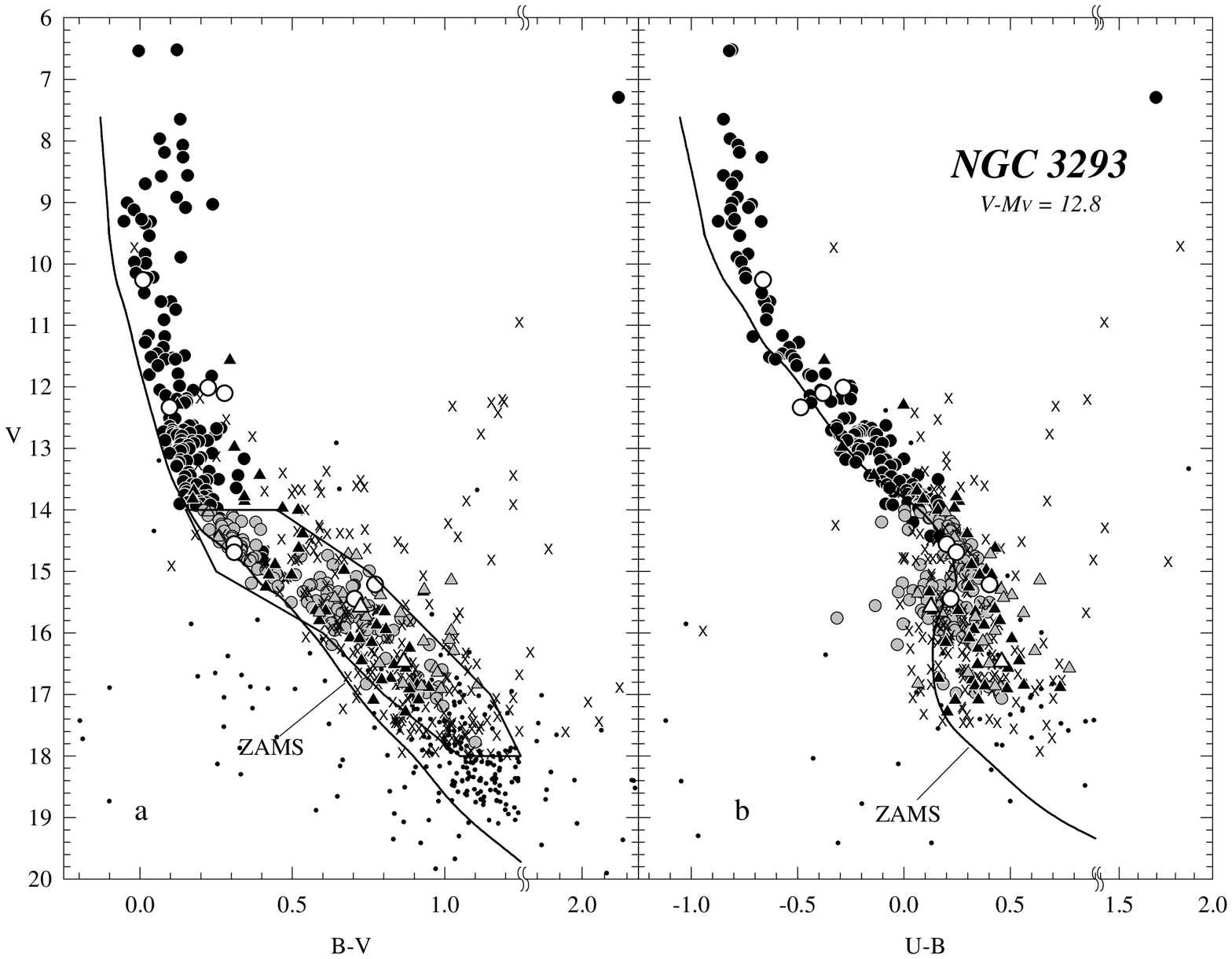}
   \includegraphics[width=14cm]{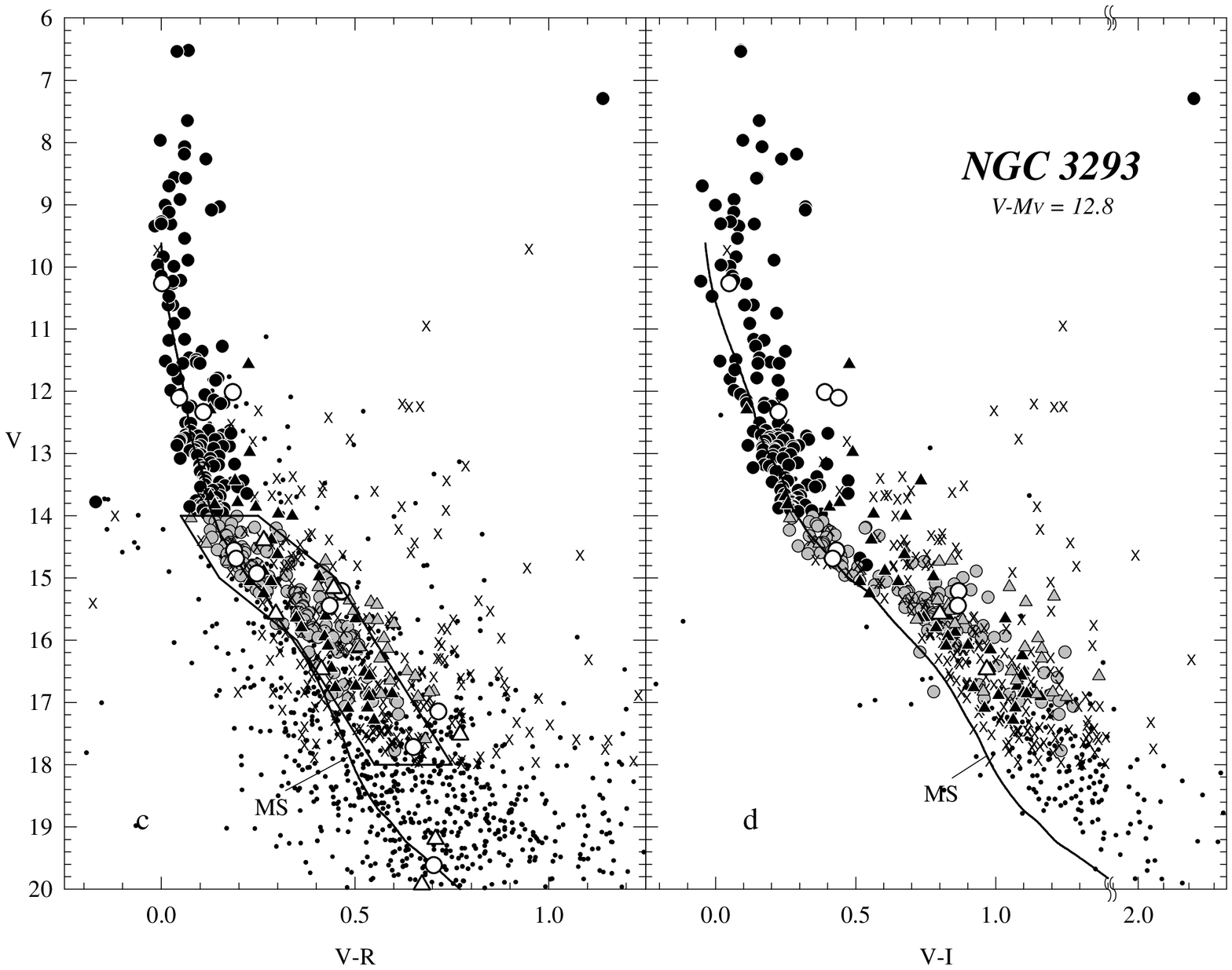}
      \caption{{}Colour--magnitude diagrams (CMDs). Symbols as in Fig 2a. Solid
      line in {\bf a} and {\bf b} is the Schmidt--Kaler's (1982) ZAMS and the
      Cousins' (1978) intrinsic line for luminosity class V stars in {\bf c} and
      {\bf d} fitted to the apparent distance modulus $V-M_{V} = 12.8$ ($V-M_{V}
      = V_{0}-M_{V}+R~(E_{B-V})_{f}$, see sections 4.1 and 4.2). Closed curves
      in {\bf a} and {\bf c} indicate limits adopted for faint stars membership
      (see section 3.2.2).}
      \label{Fig03}
\end{figure*}

$\hspace{0.5cm}$
For stars brighter than $V \approx 14$, the method described above was combined
with spectroscopic data (whenever possible) to classify stars as likely members 
(lm1) and probable members (pm1), which are indicated with black filled symbols 
in our figures. The following main features can be outlined from inspection of 
the cluster photometric diagrams:
\begin{itemize}
\item The TCD (Fig. 2a) shows, down to $B-V \approx 0.3$, a well recognisable 
      blue and scattered main sequence composed by stars with spectral types 
      earlier than A0 mostly included inside reddening values $0.20 < E_{B-V} 
      < 0.45$ (see section 4.1).
\item The four CMDs (Fig. 3), in turn, show a clear and well populated 
      upper main sequence slightly widened because of differential reddening. 
      The brightest stars of the cluster main sequence are mostly placed above 
      the ZAMS (Schmidt--Kaler 1982) while stars with $V > 13$ are on the ZAMS.
\item None of the diagrams show evidences of strong contamination of field 
      stars among bright members. Actually, most of field stars start mixing 
      with cluster members downwards $V \approx 12$ becoming an important
      obstacle to analyse the faint part of the cluster.

\end{itemize}

\subsubsection{Faint cluster members}

$\hspace{0.5cm}$
Determining memberships for stars with $V > 14$ requires a different procedure. 
If $V$ vs. $B-V$ and/or $V$ vs. $V-R$ CMDs are available for the 
$''cluster~region''$ and a $''field~region''$, we can subdivide them into a grid 
of boxes ($\Delta V = 1 \times \Delta (B-V) = 0.1$ and $\Delta V = 1 \times 
\Delta (V-R) = 0.1$) and build two--dimensional histograms for each CMD. 
Subtracting the respective two--dimensional histograms (this mean 
$''cluster~region''$ - $''field~region''$), we can remove (statistically 
speaking) the contamination produced by field stars onto the CMDs of the 
$''cluster~region''$. This way, count left on the resulting two--dimensional 
histogram define the locus occupied by cluster member stars on the CMDs (Chen et 
al. 1998). Naturally, the reliability of the number of members obtained and the 
locus they occupy, depend on two factors: a) the adopted $''field~region''$ must 
be representative of the field star distribution over the $''cluster~region''$ 
and b) the $''cluster~region''$ must include the whole extension of the cluster 
to account for mass segregation. Actually, the true cluster size will strongly 
depend on the extension of the segregation process outward it. If the 
$''field~region''$ is too close to the cluster, it may contain less massive 
segregated members; if it is far from the cluster it may not represent the true 
field star distribution across it. Finally, dust clouds and emission nebulae 
(very frequent in young clusters near the galactic plane) along with 
differential reddening lead to wrong estimations of the field star distribution 
too (Mermilliod 1976; Prisinzano et al. 2001; see Forbes 1996 for details). \\

In the remaining of this section all procedures will be applied only to stars 
in the range $14 < V < 18$ with photometric errors $< 0.1$. A first point to 
treat is whether comparison fields around NGC 3293 do show any strong 
differences produced by random stellar fluctuations and/or the type of stellar 
data. To ease the analysis we divided our data set into four groups:
\begin{itemize}
\item $Sample~I$: $V$, $B-V$ and $V-R$ data from stars placed in the cluster 
      area, up to a $4'\!.1$ radius. This sample is the $''cluster~region''$ and 
      contains cluster stars plus field stars.
\item $Sample~II$: $V$, $B-V$ and $V-R$ data for stars placed between 
      $4'\!.1$ radius and the limits of the grey line squared frames shown in 
      Fig. 1.
\item $Sample~III$: $V$ and $B-V$ data from stars placed in the comparison 
      field (see section 2) $20'$ north of the cluster.
\item $Sample~IV$: $V$ and $V-R$ data for stars from the grey line squared
      frames outwards.
\end{itemize}

\begin{figure*}
   \centering
   \includegraphics[width=16cm]{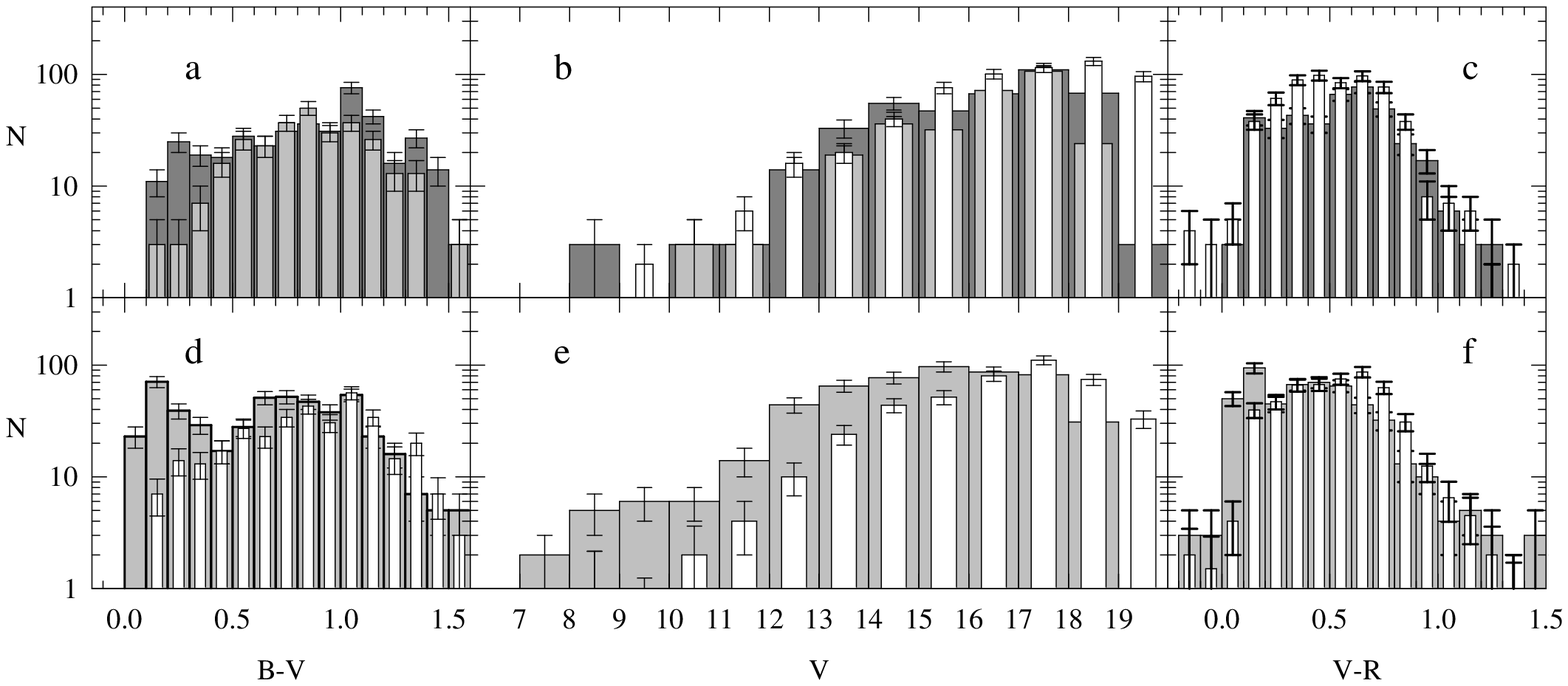}
      \caption{Marginal distributions of the two--dimensional histograms of the
       different samples taken from our data (see section 3.2.2). Upper panels, 
       {\bf a},{\bf b} and {\bf c}, represent what we call $''field~regions''$: 
       dark--grey distributions belong to $Sample~II$, grey ones to $Sample~III$ 
       and white ones to $Sample~IV$. In {\bf d}, {\bf e} and {\bf f}, grey 
       distributions belong to $Sample~I$ ($''cluster~region''$ containing 
       cluster plus field stars) while the white ones stand now for the average 
       of samples $II$, $III$ and $IV$ (our best representation of the 
       $''field~region''$}
      \label{Fig04}
\end{figure*}

We warn that samples $II$, $III$ and $IV$ are different representations of what 
we called $''field~region''$. As they cover different areas they were
adequately scaled to the $''cluster~region''$ area to built the two-dimensional 
histograms of each sample. Any strong spatial variation among the field samples 
should be revealed by the marginal distributions of the two--dimensional 
histograms of each sample. A brief inspection of them, shown in Fig. 4, let us 
say:
\begin{itemize}
\item Figures 4b and 4e show a same degree of completeness of the photometry in 
      the $''cluster~region''$ and the field samples down to $V \approx18$.
\item Figs. 4a-c, show that different field samples have not only the same shape 
      (down to $V \approx 18$) but also the same number of stars 
      approximately. Aside from uncertainties produced by small number 
      statistics, the actual true distribution of field stars over the cluster 
      surface should not differ too much from these three.
\end{itemize}

To quantify the last issue, we applied a Kolmogorov--Smirnoff test to the field 
distributions confirming that, for $V < 18$, the three field samples are 
similar at the level $\alpha = 0.05$. Thus, the average of their respective 
two--dimensional histograms ($V$ vs. $B-V$ and/or $V$ vs. $V-R$) yields the best 
representation of the $''field~region''$ across NGC 3293. The corresponding 
marginal distributions are presented as white histograms in Figs. 4d-f). \\

Another point to analyse is whether NGC 3293 has undergone mass segregation
and shows a core/halo structure, as suggested by HM82. Although in section 3.1 
we did not find evidences of any halo around NGC 3293, we compared the ratios of 
stellar densities found in the $''cluster~region''$ ($\rho_{IN}$) and outside 
it ($\rho_{OUT}$) as a function of $V$. If an appreciable amount of less massive 
(faint) stars were located outside the cluster boundaries due to an active mass 
segregation process then the $\rho_{OUT}/\rho_{IN}$ ratio should show that. As 
expected, the plots of Fig. 5 do not indicate any appreciable star over excess 
outside the cluster area, but just a very slow increasing at very faint 
magnitudes (at the level where the completeness starts being questionable). We 
conclude that the cluster has not undergone mass segregation and its limits from 
section 3.1 are fully reliable. \\

We proceeded then to subtract the average two--dimensional histogram of the 
$''field~region''$ from the one of the $''cluster~region''$ obtaining the 
contamination--free $V$ vs. $B-V$ and $V$ vs. $V-R$ distributions. They
revealed then the presence of stellar bands above the ZAMS with confident lower 
and upper limits. Such limits were used on the CMDs (see Figs. 3a and 3c) to 
define which stars are (from a statistical point of view) cluster members 
following the next criteria:
\begin{itemize}
\item A star is a cluster likely member (lm2) if it is found inside the 
      cluster boundaries defined in section 3.1 and is simultaneously included
      inside the band limits in the two CMDs ($V$ vs. $B-V$ and $V$ vs. $V-R$).
\item A star is a probable member (pm2) if, apart from being spatially well 
      located, it fits, at least, in any of high density zones defined in the 
      CMDs and shows slight departures in the other.
\end{itemize}

Finally, the resulting sample was filtered (following a random distribution) in 
order to obtain final CMDs with an amount of stars in each colour--magnitude box 
similar to that obtained after the subtraction of the two--dimensional 
histograms. The adopted memberships are shown with grey filled symbols in the 
photometric diagrams of our figures. Notice in advance member stars with 
$V > 15.5$ mainly placed above the ZAMS line in the $V$ vs. $B-V$ and $V$ vs. 
$V-I$ diagrams (we will return to this point in subsection 4.3 and section 
6). \\

\begin{figure}[b]
   \centering
   \includegraphics[width=8cm]{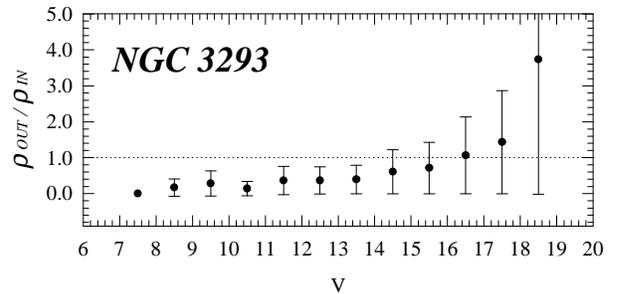}
      \caption{{}$\rho_{OUT}/\rho_{IN}$ vs. $V$ diagram. Poisson statistic error
      in each magnitude are also indicated.}
      \label{Fig05}
\end{figure}

\section{Cluster Parameters}

\subsection{Corrected colours and magnitudes of cluster members}

$\hspace{0.5cm}$
For the estimation of the cluster mean colour excesses, we used firstly the 14 
stars with known spectral classification and luminosity class IV--V (see Table 
3) using the Schmidt--Kaler (1982) relations of spectral types and intrinsic 
colours (Be stars were excluded). The plot of excesses is shown in Fig. 6a 
together with the standard $E_{U-B}/E_{B-V}$ relation. The fitting is poor 
probably because of colour anomalies of the many variable stars present in 
the sample. Mean value excesses of these stars were $E_{B-V} = 0.29 \pm 
0.06~(s.d.)$ and $E_{U-B} = 0.23 \pm 0.08~(s.d.)$. Secondly, we applied the well 
known relations $E_{U-B}/E_{B-V} = 0.72 + 0.05~E_{B-V}$ and $(U-B)_{0} = 
3.69~(B-V)_{0} + 0.03$ to get the intrinsic colours of those stars with $V < 14$ 
adopted as likely members and without spectral classification (mostly located at 
$0.20 < E_{B-V} < 0.45$, see Fig. 2a). When we included this last group of 
stars in the mean colour excesses computations, the obtained values were 
$E_{B-V} = 0.29 \pm 0.06~(s.d.)$ and $E_{U-B} = 0.21 \pm 0.05~(s.d.)$, almost 
identical to the obtained firstly. We de-reddened cluster members with $V < 14$
(except star \#3) using individual excesses, meanwhile other cluster members 
and star \#3 were de-reddened using these lasts mean excess values. The 
estimated foreground colour excesses of NGC 3293 were $(E_{B-V})_{f} = 0.20$ and 
$(E_{U-B})_{f}=0.15$, slightly lower than the typical ones for southern Carina 
(Tr 15, Carraro 2002; Tr 14/15/16, Tapia et al. 2002). \\

The next step was to know the absorption law valid in NGC 3293 given by 
$R = A_{V}/E_{B-V}$. Galactic regions with normal absorption have a mean of 
$R = 3.1-3.2$ although larger $R$--values are especially found in regions of 
recent star formation. To compute the local $R$--value we obtained individual 
$E_{B-V}$, $E_{V-R}$ and $E_{V-I}$ excesses through the $(B-V)_{0}$ with 
$(V-R)_{0}$ and $(V-I)_{0}$ relations for stars without spectral classification 
(circles in Fig. 6b-c) and the relation between MK types and $(V-R)_{0}$ and 
$(V-I)_{0}$ for stars with spectral types (squares in Fig. 6b-c), both from 
Cousins (1978). $E_{V-R}/E_{B-V}$ and $E_{V-I}/E_{B-V}$ ratios depend on the 
$R_V$--values as when $E_{V-R}/E_{B-V} = 0.57$ and $E_{V-I}/E_{B-V} = 1.244$ 
the interstellar material is normal (V\'azquez et al. 1995; Dean et al. 1978). 
The plot of 140 stars with $E_{B-V}$, $E_{V-R}$ and $E_{V-I}$ are depicted in 
Figs. 6b and 6c showing an excellent agreement with the typical reddening 
relations. The mean ratios found are $E_{V-R}/E_{B-V} = 0.54 \pm 0.15~(s.d.)$ 
and $E_{V-I}/E_{B-V} = 1.26 \pm 0.18~(s.d.)$, indicative of a normal reddening 
law for which is $R = 3.1$. The foreground excesses we found were 
$(E_{V-I})_{f} = 0.25$ and $(E_{V-R})_{f} = 0.11$. Finally we mention that FM80 
claim that $R$ can raise up to $3.5$. That is marginally probable seeing the 
plots of Fig. 6b-c, although the scatter around the mean lines may be mainly 
caused by circumstellar envelopes, variability and/or binarity instead of 
anomalies in the absorption. Therefore, we adopted $R = 3.1$ to obtain corrected 
magnitudes as $V_{0} = V - 3.1~E_{B-V}$. \\

   \begin{table*}
      \caption[]{Main characteristics of bright stars in NGC 3293}
      \fontsize{8} {10pt}\selectfont
      \begin{tabular}{rrll@{-}l@{-}lccccc}
      \hline
         $\#$ & $\#_{FM}$ & \multicolumn{1}{c}{$ST$} & \multicolumn{3}{c}{Remarks} & $E_{B-V}$ & $E_{U-B}$ & $E_{V-R}$ & $E_{V-I}$ & $V_{o}-M_{V}$ \\
      \hline
       1 &  3~~ & B0..5 Iab & V513 Car~~ & ~~Hip-Tyc~~ &                               &     0.34 &     0.22 &     0.18 &     0.38 &    11.86 \\
       2 &  4~~ & B0 Ib     &            & ~~Hip-Tyc   &                               &     0.24 &     0.25 &     0.15 &     0.38 &    11.91 \\
       3 & 21~~ & M1.5 Iab  & V361 Car   & ~~IRAS      &                               &          &          &          &          &    12.00 \\
       4 & 22~~ & B1 II     &            &             & ~~s.d.s.                      &     0.37 &     0.14 &     0.18 &     0.45 &    11.90 \\
       5 & 20~~ & B1 III    & V439 Car   &             &                               &     0.33 &     0.15 &     0.11 &     0.39 &    11.36 \\
       6 & 25~~ & B1 III    &            &             &                               &     0.40 &     0.19 &     0.17 &     0.46 &    11.23 \\
       7 &  6~~ & B0.5 III  &            &             &                               &     0.36 &     0.27 &     0.17 &     0.58 &    11.84 \\
       8 &  8~~ & B0.5 III  &            & ~~Hip-Tyc   &                               &     0.42 &     0.37 &     0.23 &     0.53 &    11.72 \\
       9 & 26~~ & B1 III    & V379 Car   &             & ~~Be$^{(3)}$ - $\beta$Ceph    &     0.42 &     0.12 &     0.14 &     0.44 &    11.67 \\
      10 &  7~~ & B1 III    &            &             & ~~b.s.$^{(1)}$                &     0.33 &     0.19 &     0.17 &     0.44 &    11.95 \\
      11 & 16~~ & B1 IV     & V403 Car   &             & ~~$\beta$Ceph                 &     0.28 &     0.15 &     0.13 &     0.24 &    11.44 \\
      12 & 27~~ & B0.5 III  & V380 Car   &             & ~~$\beta$Ceph                 &     0.40 &     0.26 &     0.16 &     0.36 &    12.44 \\
      13 &  5~~ & B1 III    & V381 Car   &             & ~~$\beta$Ceph - b.s.$^{(1)}$  &     0.22 &     0.16 &     0.12 &     0.29 &    12.73 \\
      14 & 19~~ & B1 III    &            &             & ~~s.d.s.                      &     0.50 &     0.25 &     0.26 &     0.61 &    11.89 \\
      15 &  2~~ & B1 III    &            & ~~Hyp-Tyc   &                               &     0.41 &     0.24 &     0.24 &     0.61 &    12.22 \\
      17 & 18~~ & B1 V      & V406 Car   &             & ~~$\beta$Ceph                 &     0.27 &     0.15 &     0.11 &     0.34 &    11.65 \\
      18 & 14~~ & B0.5 V    & V405 Car   &             & ~~$\beta$Ceph                 &     0.23 &     0.14 &     0.11 &     0.31 &    12.20 \\
      19 & 24~~ & B1 III    & V378 Car   &             & ~~$\beta$Ceph                 &     0.29 &     0.30 &     0.13 &     0.43 &    12.80 \\
      20 & 23~~ & B1 III    & V404 Car   &             & ~~$\beta$Ceph - s.d.s.        &     0.28 &     0.16 &     0.09 &     0.37 &    12.89 \\
      21 & 10~~ & B1 V      & V401 Car   &             & ~~$\beta$Ceph                 &     0.29 &     0.18 &     0.17 &     0.37 &    11.84 \\
      22 & 43~~ & K5        &            & ~~Hip-Tyc   & ~~r.fg.s.                     &          &          &          &          &          \\
      23 & 42~~ & A0        &            & ~~Hip-Tyc   &                               &          &          &          &          &          \\
      25 & 12~~ & B1 V      & V402 Car   &             & ~~Be$^{(2)}$ - b.s.$^{(1)}$   &     0.39 &     0.16 &     0.18 &     0.50 &    12.77 \\
      27 &  9~~ & B2 V      &            & ~~Hip-Tyc   & ~~b.s.$^{(1)}$                &     0.26 &     0.07 &     0.13 &     0.31 &    11.59 \\
      28 & 13~~ & B1.5 V    &            &             &                               &     0.24 &     0.15 &     0.10 &     0.34 &    12.22 \\
      30 & 15~~ & B1 V      &            &             & ~~s.d.s.                      &     0.28 &     0.21 &     0.14 &     0.24 &    12.55 \\
      31 & 28~~ & B1 V      &            &             & ~~$H_{\alpha}$ - b.s.$^{(1)}$ &     0.27 &     0.29 &     0.11 &     0.34 &    12.63 \\
      33 & 17~~ & B2.5 V    & V440 Car   &             &                               &     0.23 &     0.11 &     0.11 &     0.22 &    11.75 \\
      34 & 33~~ & B2 V      &            &             &                               &     0.34 &     0.21 &     0.13 &     0.40 &    11.95 \\
      35 & 29~~ & B2 V      &            &             &                               &     0.31 &     0.18 &     0.12 &     0.36 &    12.06 \\
      38 &  -~~ & B8        &            & ~~Hip-Tyc   & ~~b.bg.s.                     &          &          &          &          &          \\
      47 & 31~~ & B2 V      &            &             &                               &     0.28 &     0.21 &     0.11 &     0.28 &    13.06 \\
      48 &  -~~ &           &            & ~~Hip-Tyc   &                               &          &          &          &          &          \\      
      96 & 30~~ & B5 V      &            &             &                               &     0.27 &     0.40 &     0.15 &     0.29 &    13.01 \\
     115 & 35~~ & B7 V      &            &             &                               &     0.27 &     0.17 &     0.09 &     0.29 &    12.64 \\
     116 & 34~~ & B8 V      &            &             &                               &     0.19 &     0.07 &     0.08 &     0.22 &    12.47 \\
     120 & 32~~ & B8 V      &            &             & ~~Be$^{(1)}$                  &     0.19 &     0.21 &     0.13 &     0.27 &    12.50 \\
  	\hline
    \end{tabular}\\
    \begin{tabular}{lll}
       {\bf Remarks:}                    &                                                    & {\bf Notes:}                                                        \\
        b.s. = binary star               & $\beta$Ceph = beta cepheid star                    & - $\#_{FM}$ indicates numbering from FM80.                          \\
        s.d.s. = star in double system   & $H_{\alpha}$ = likely $H_{\alpha}$ emission star   & - Spectral classification was taken from TGHH80 and FM80.           \\
        r.fg.s. = red foreground star    & Be = Be star                                       & - $V_{0}-M_{V}$ column contains spectrophotometric distance moduli. \\
        b.bg.s. = blue background star   & IRAS = Source IRAS 10338-5756                      & - Remarks information is mainly from $SIMBAD$ database.             \\ 
    \end{tabular}\\
    \begin{tabular}{l}
    \begin{minipage}{16cm} 
       Hip-Tyc = Star with paralax and proper motion measurements \\
       (see Table 1 for HIP/TYC identification)\\
       $(1)$ = Feast (1958); $(2)$ = Schild (1970); $(3)$ = Shobbrook (1980)\\
    \end{minipage}\\
    \end{tabular}
   \end{table*}
   
\subsection{Cluster distance and size}

$\hspace{0.5cm}$
The distance of NGC 3293 was derived superposing the Schmidt--Kaler's (1982) 
ZAMS onto the reddening--free CMD. The best ZAMS fitting was achieved for a 
distance modulus $V_{0} - M_{V} = 12.2 \pm 0.2$ (error from eye--inspection). 
We also applied the spectroscopic parallax method to 33 stars to get the cluster 
distance modulus (using the relation of spectral types and $M_{V}$ from 
Schmidt--Kaler 1982) that yielded $12.1 \pm 0.5$. If only 17 stars of luminosity 
class IV--V are used, the distance modulus turns out to be $12.3 \pm 0.5$ (see 
last column in Table 3). The large dispersions of the moduli associated to the 
spectroscopic parallax method may be intrinsic for early type stars (Conti \& 
Underhill 1988) or produced by a high percentage of variable and binary stars 
(Feast, 1958). Our distance modulus is a bit larger than the ones given by Feast 
(1958), $12.08 \pm 0.1$, FM80, $12.1 \pm 0.15$, TGHH80 $11.99 \pm 0.13$, 
Shobbrook (1983), $11.95 \pm 0.1$ and lesser than the one from Balona \& 
Crampton (1974), $12.32 \pm 0.09$. Notwithstanding, as at $1 \sigma$, all of 
them are almost coincident, we adopted $V_{0} - M_{V} = 12.2 \pm 0.2$ which 
yields a cluster distance $d = 2750 \pm 250~pc$. The absolute magnitude scale 
$M_{V}$ shown in Fig. 7 was set with this mean value. \\

\begin{figure}
   \centering
   \includegraphics[width=7.5cm]{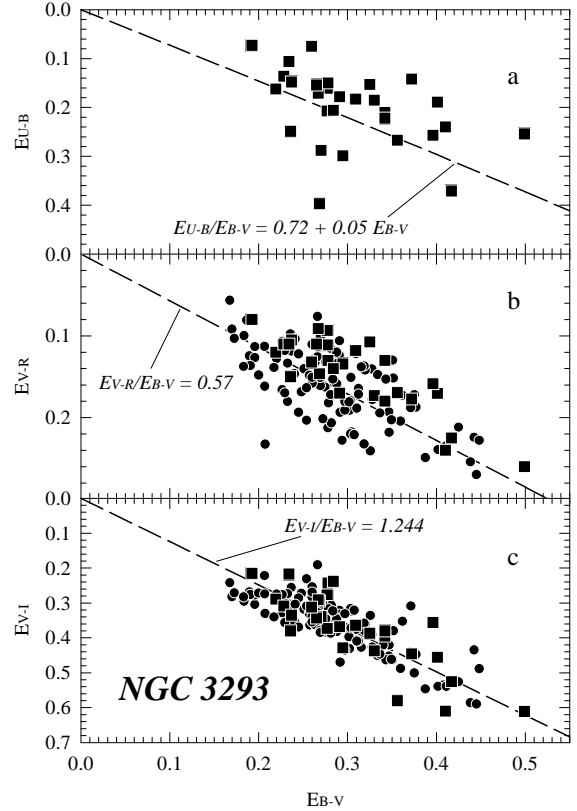}
      \caption{{} $E_{U-B}$, $E_{V-R}$ and $E_{V-I}$ vs. $E_{B-V}$ diagrams for
      likely cluster members with $V < 14$ except star \#3 and Be stars.
      Squares are stars with available spectral classification and dashed
      curves indicate normal relations.}
      \label{Fig06}
\end{figure}

When the above distance modulus is combined with the angular radius obtained in 
section 3.1 it yields a linear cluster radius $3.3 \pm 0.3~pc$, comparable to 
the sizes of other studied young open clusters as Cr 272, $4.0~pc$; NGC 6231, 
$4.1~pc$, and H-M 1, $2.9~pc$ (from V\'{a}zquez et al. 1997; Baume et al. 1999 
and V\'{a}zquez \& Baume 2001 respectively). This result however disagrees with 
the Janes et al. (1988) result that most of young clusters have diameters less 
than $5~pc$. \\

\begin{figure*}
   \centering
   \includegraphics[width=14cm]{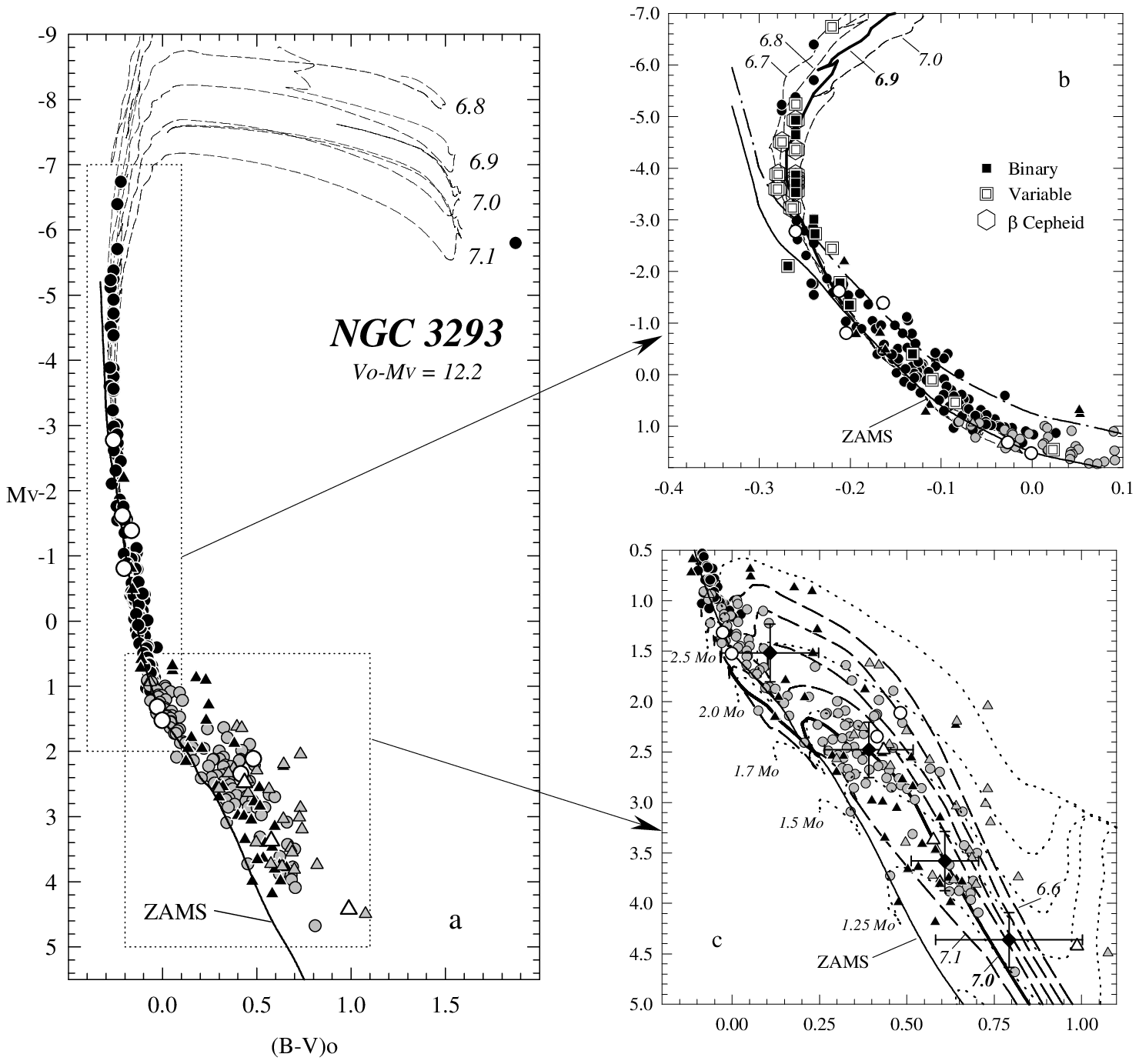}
      \caption{{\bf a)} The entire $M_{V}$ vs. $(B-V)_{0}$ diagram. Symbols 
      as in Fig. 2a. Solid line is the Schmidt--Kaler's (1982) ZAMS. Dashed 
      lines are the isochrones from Girardi et al. (2000). Numbers give the 
      $\log{age}$. {\bf b)} An enlargement of the upper main sequence showing 
      the binary envelope (dotted -- dashed line) $0.75^{m}$ above the ZAMS. The 
      thick line is the isochrone of $8~Myr$. Numbers as in {\bf a}. {\bf c)} An 
      enlargement showing the faint sequence of the cluster. Dotted and dashed 
      lines represent the tracks and isochrones (Bernasconi \& Maeder 1996). 
      Numbers indicate the masses of each track and the $\log(age)$. The 
      $10~Myr$ isochrone is shown as a thick line. Mean values and dispersions 
      of PMS stars grouped at intervals $\Delta M_{V} = 1$ are also indicated}
     \label{Fig07}
\end{figure*}
 
\subsection{Cluster age} 

\subsubsection{Nuclear age} 

$\hspace{0.5cm}$
Regarding the nuclear age of NGC 3293, the isochrones derived from Girardi et 
al. (2000) evolutionary models (computed with solar metallicity, mass loss and 
overshooting) are shown in Fig. 7a-b superposed to the cluster upper main 
sequence. Because an important fraction of bright members were de--reddened 
using their spectral types some scatter arises that precludes a single 
isochrone to fit the bright part of NGC 3293; binaries and fast rotators are 
likely to be another source of scatter in upper CMD. The envelope of binaries 
$0.75^{m}$ above the ZAMS encloses pretty well most of binary stars in the upper 
main sequence. We draw the attention on the red super--giant (star \#3) which is 
not included by any isochrone (Fig. 7a), but it lies close to the isochrones of 
$10-12.6~Myr$. As many other red--supergiant stars, this star is not contained 
by any isochrone, a frequent effect already reported in Meynet et al. (1993). \\

From stars with $M_{V} < -4$, the nuclear age of NGC 3293 goes from $6.5$ to 
$10.0~Myr$ with a probable mean age of $8~Myr$. The cluster is thus older than 
in previous investigations, $5 \pm 2~Myr$ by TGHH80 and $6 \pm 2~Myr$ from HM82. 
Such differences may have their origin in different (smaller) distance moduli 
estimates and also in the use of different isochrone sets. \\
 
\subsubsection{Contraction age} 

$\hspace{0.5cm}$ 
As indicated in section 3.2.2, a relevant feature emerging from the cluster 
lower sequence (Fig. 7c) is that the left envelope of NGC 3293 does not follow 
the shape of the ZAMS. The lower cluster sequence shows a bend at $M_{V} \approx 
+2$, lower of which it lies $1^{m}$ above the ZAMS approximately 
constituting a parallel sequence that confirms the earlier assertion of HM82 
that faint stars are mostly above the ZAMS. \\
 
As contamination of field interlopers has been already removed, the stars in 
this parallel sequence have to be interpreted as stars in contraction phase 
towards the ZAMS. The strong $M_{V}$ scatter at constant colour is a normal 
feature associated to the age spread (Iben \& Talbot 1966) among these type of 
objects (cf. Fig. 3 in Preibisch \& Zinnecker 1999). However, the contrast 
between the wide PMS band and the sharp upper main sequence in Fig. 7c is 
produced primarily by: a) differential reddening, that was not removed from the 
main sequence faint stars (they were de--reddened using the mean excess values 
derived in section 4.1) and, b) several intrinsic factors detailed below. In 
overlaying in Fig. 7c the accretion evolutionary models developed by Bernasconi 
\& Maeder (1996) we find that the PMS objects have masses ranging from $1$ to 
$2.5 \cal M_{\sun}$ and ages from $6$ to $12~Myr$, a PMS age range close to 
$\sim 5~Myr$ found recently in the other Carina clusters Tr 14, Tr 15 and Tr 16 
by Tapia et al. (2002) and confirmed in Tr 15 by Carraro (2002). \\

We do not find a single isochrone that fits the whole lower sequence; indeed PMS 
stars tend to cross several of them. This lack of alignment led Iben \& Talbot 
(1966) to reject out the hypothesis of coeval star formation in the clusters NGC 
2264 and NGC 6530. Recently, the same was done in NGC 2264 by Flaccomio et al. 
(1999). Notwithstanding, departures from a single isochrone produced by 
differential reddening, binarity/multiplicity, random distribution of accretion 
discs around single and multiple systems, photometric errors and physical 
uncertainties in the evolutionary models are expected to happen. The influence 
of differential reddening is hard to estimate as it affects in different manners 
different stars. As for binarity, it is very frequent not only among normal less 
massive stars ($> 50\%$, Bessell \& Stringfellow 1993; Preibisch \& Zinnecker 
1999) but also among PMS stars (Hartigan et al. 1994). Binarity raises the stars 
above a reference line (e.g. the ZAMS) in a way that depends on the mass ratio 
of the binaries. As for the accretion stellar discs, their occultation and 
emission, and also the random orientation of their angles, coupled with a 
possible range of accretion rates (Kenyon \& Hartmann 1990) introduce more 
scatter among the PMS population. \\

To smooth these effects, the mean of the stellar distribution in Fig. 7c was 
computed. Interestingly, the mean follows closely the isochrone of $10~Myr$ 
shown by a thick line, except at $M_{V} = +1.5$. This is a magnitude point where 
very often (Rachford \& Canterna 2000, Phelps \& Janes 1993; de Bruijne et al. 
2000) a gap appears in many open clusters so that the statistics here may be 
irrelevant. Therefore we adopt $10~Myr$ as the mean contraction age of the PMS 
population in NGC 3293. \\
 
\subsubsection{Star formation rate} 

$\hspace{0.5cm}$ 
Does the fact of finding a $6~Myr$ age spread among PMS objects indicate the 
star formation process is not coeval? We examined the point computing individual 
present--day masses of PMS stars and giving them the corresponding ages 
interpolating among Bernasconi \& Maeder (1996) models. Following Iben \& 
Talbot (1966) and Adams et al. (1983) definitions and procedures, we estimated 
the star formation rate (SFR) in the cluster for three different stellar masses 
bins. The results, shown in Fig. 8, suggest no dependence of the SFR with 
stellar masses when allowing for the uncertainties involved in that kind of 
computation (see Stahler 1985). Actually it seems to confirm that the stellar 
formation in NGC 3293 took place in a very short period of time, mostly 
between $7$ and $10~Myr$ ago. \\
 
\begin{figure}
   \centering
   \includegraphics[width=7.5cm]{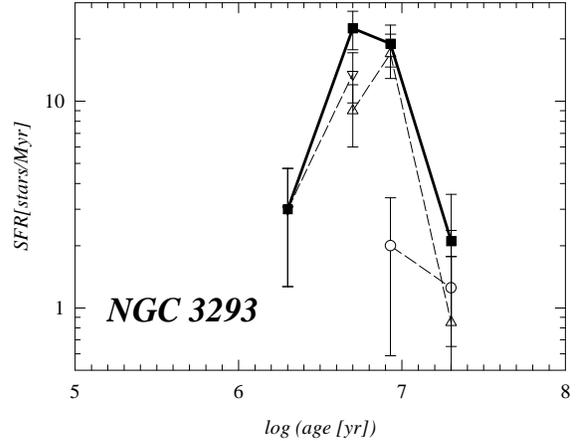}
      \caption{Star formation rate (SFR) in NGC 3293 from Bernasconi \& Maeder
      (1996) evolutionary models. Circles, upside triangles and downside
      triangles indicate SFR values for stars with masses in the intervals
      $1.0-1.5$, $1.5-2.0$ and $2.0-2.5 \cal M_{\sun}$ respectively. Thick line 
      (black squares) represents the total SFR. Error bars are from Poisson
      statistic.}
      \label{Fig08}
\end{figure}
 
\subsubsection{Nuclear and contraction ages}

$\hspace{0.5cm}$
The mean contraction age of the entire PMS population becomes comparable to the 
mean nuclear age deduced from stars with \( M_V < -4 \) since there is no 
substantial difference between the $8~Myr$ and $10~Myr$ ages of evolved and PMS 
stars respectively. Most important is that the early original age discrepancy of 
$20~Myr$ found by HM82 is now strongly reduced. On the other hand, as mentioned 
above, companion stars (Preibisch \& Zinnecker 1999) overestimate the luminosity 
of a given star (e.g., underestimating the luminosity of an M0 star by a factor 
of 2 reduces its age by a factor of 4 and for a factor of 2 for a G5 star). 
Also, the net effect of accretion stellar discs, as was mentioned above, 
introduce scatter in luminosities that produce uncertainties of a factor of 
$2 - 3$ in the age of an individual object (Kenyon \& Hartmann 1990). We are 
confident therefore, that the $2~Myr$ difference found between the mean 
contraction age and the nuclear age, results irrelevant in view of the mentioned 
effects. \\
 
Moreover, the $0.2^{m}$ error of the distance modulus can reduce or increase the 
difference between both ages without affecting the nuclear age which remains 
almost invariable. Besides, the ``turn on'' point position (see Fig. 7c) is not 
very accurately determined but, allowing for the above uncertainties, it can 
be placed at $M_{V} \approx +2$ corresponding to a stellar mass 
$\sim 2 \cal M_{\sun}$ with a contraction age of $7.9~Myr$ close to the nuclear 
age. \\

\subsection{Cluster LF and IMF}

\subsubsection{The Luminosity Function} 

$\hspace{0.5cm}$ 
The cluster LF gives the fraction of stars in each absolute magnitude bin of 
size $\Delta M_{V} = 1$. The grey histogram in Fig. 9 shows the LF of NGC 3293 
where known cluster binaries (see Table 3) were all corrected by $0.75$ and two 
stars were counted instead of one. For comparison purposes we included also the 
cluster LF earlier determined by HM82 and the cluster overall LF found by Phelps 
\& Janes (1993). Except by the strong dip present at $M_{V} = -2$ in the HM82 
LF, there is no important differences among the three LFs down to $M_{V} \approx 
+1$. From $M_{V} \approx +1$, HM82 LF differs from ours as the sharp dip at 
$M_{V} \approx +3$ is not revealed by our data. Somehow, that confirms the 
suspicion of HM82 that the dip they found is an artifact produced by a 
combination of incompleteness and eye-estimate of field faint star magnitudes. 
Interesting enough, our LF as well as the HM82's show an apparent low number of 
stars for $M_{V} > 0$ when comparing with the Phelps \& Janes estimation: ours, 
in particular, shows a flattening in the range $0 < M_{V} < +3 - +4$. At this 
point it is necessary to notice that all stars below $M_{V} \approx +2$ are PMS 
objects as we have already stated and their luminosities do not represent the 
luminosities they will have in the ZAMS; depending on the star location, the 
present star luminosity and the final in the ZAMS will differ for more than 
$1^{m}$ changing the LF shape. Therefore, the cluster LF is only comparable to 
another from main sequence objects (Wilner \& Lada, 1991) in the range 
$-7 < M_{V} < +2$. \\

\begin{figure}[b]
   \centering
   \includegraphics[width=8cm]{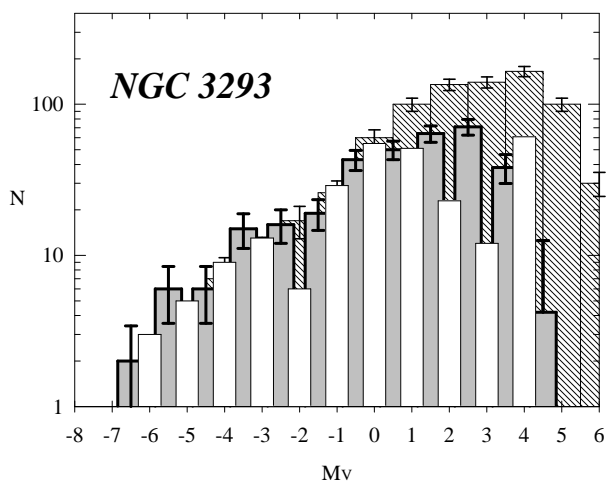}
      \caption{Luminosity Function of NGC 3293 (grey histogram). Error bars 
       are from Poisson statistic and the sustraction process. The white 
       histogram is the LF from HM82 and the hatched one is the combined LF of 
       several young open clusters from Phelps \& Janes (1993).}
      \label{Fig09}
\end{figure}

\subsubsection{The Initial Mass Function} 

$\hspace{0.5cm}$
Stellar masses were derived using an interpolation process (Baume et al 1994) 
converting the $M_V$, $(B-V)_0$ and $(U-B)_0$ values into $\log{L}$ and $\log 
{T_{eff}}$ first, where a mass is assigned in this theoretical plane. The code 
uses bolometric corrections from Schmidt-Kaler and interpolates among 
evolutionary tracks reconstructing the path of a given star backwards to its 
original point in the ZAMS. Bernasconi \& Maeder (1996) tracks were used for 
masses $< 2.5 \cal M_{\sun}$ and Girardi et al. (2000) tracks for stars 
above that mass limit. Appropriate mass bins were adopted to distribute stellar 
masses as shown in Table 4 and the mass points are depicted on top of Fig. 10 
with the $\sqrt{N}$ count bars. To get the slope of the cluster IMF ($x$) we 
used a weighted least squares method applied to different mass ranges whose 
results are included at bottom of Table 4. The lowest mass bin has been excluded 
of any calculations as it may be affected by effects described below. The 
fitting of the most massive stars gives a steep slope with a high error value
($x = 1.6 \pm 0.5$), however the portion that includes low and intermediate mass 
stars yields a flatter slope ($x = 1.2 \pm 0.2$). The last value is closer to
the typical slope for field stars ($x = 1.35$) and only marginally comparable 
to the $x = 0.9$ found by HM82. \\

\begin{figure}
   \centering
   \includegraphics[width=7cm]{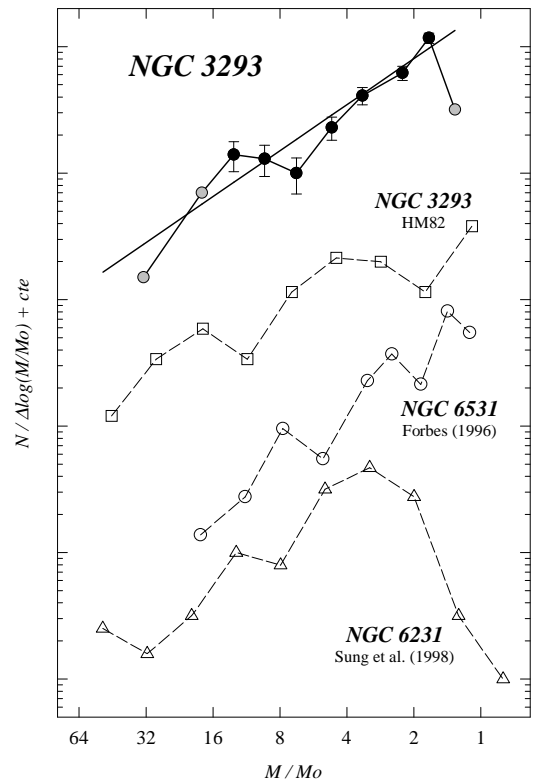}
     \caption{Initial Mass Function (IMF). The IMF obtained for NGC 3293 is
      shown at the top (error bars are from Poisson statistic). The weighted 
      least square fitting for the mass range $1.41 < \cal{M}/\cal{M_{\sun}} 
      <$ $16.0$ is also presented (points not included in the fit are indicated 
      with grey symbols). Other IMFs for comparable clusters are also shown.}
      \label{Fig10}
\end{figure}

Figure 10 also includes IMFs of cluster of similar ages and the earlier IMF of 
NGC 3293 determination made by HM82. Interesting common features are shown by 
the three clusters in the figure. There is always an increasing star number till 
a dip or flat zone appears. The mass range of the dip is not the same in the 
three clusters. While ours happens at $8 \cal M_{\sun}$, it happens at 
$10 \cal M_{\sun}$ in the HM82 IMF. Coincident with ours is the dip at 
$8 \cal M_{\sun}$  in NGC 6231 but the most strange location corresponds to the 
IMF of NGC 6531 whose dip happens at $6 \cal M_{\sun}$. The pattern is not easy 
to explain but the differences in the mass may reflect differences in the method
of assigning masses on the ZAMS according to the evolutionary models used. \\

   \begin{table}
      \centering
      \caption[]{{}The Initial Mass Function.} 
      \begin{tabular}{ccc} 
      \hline 
          $\Delta \cal M / \cal M_{\sun}$ & $\Delta M_{V}^{(a)}$ & $N^{(b)}/\Delta\log(\cal M / \cal M_{\sun})$ \\ 
      \hline 
        $22.63 - 45.25$ & $-7.0 - -6.0$ & $~~1.5 \pm ~1.2$ \\
        $16.00 - 22.63$ & $-6.0 - -4.2$ & $~~7.0 \pm ~2.6$ \\
        $11.31 - 16.00$ & $-4.2 - -3.1$ & $~14.0 \pm ~3.7$ \\
        $~8.00 - 11.31$ & $-3.1 - -2.2$ & $~13.0 \pm ~3.6$ \\
        $~5.66 - ~8.00$ & $-2.2 - -1.3$ & $~10.0 \pm ~3.2$ \\
        $~4.00 - ~5.66$ & $-1.3 - -0.5$ & $~23.0 \pm ~4.8$ \\
        $~2.83 - ~4.00$ & $-0.5 - +0.4$ & $~41.0 \pm ~6.4$ \\
        $~2.00 - ~2.83$ & $+0.4 - +1.6$ & $~62.0 \pm ~8.0$ \\
        $~1.41 - ~2.00$ & $+1.6 - +3.3$ & $117.0 \pm 10.8$ \\
        $~1.00 - ~1.41$ & $+3.3 - +5.0$ & $~31.8 \pm ~8.3$ \\
      \hline 
          $\Delta \cal M / \cal M_{\sun}$ & $x$ & \\
      \hline 
         $1.41 - 45.25$ & $-1.23 \pm 0.13$ \\
         $1.41 - 16.00$ & $-1.21 \pm 0.18$ \\           
         $8.00 - 45.25$ & $-1.63 \pm 0.52$ \\              
      \hline
      \end{tabular} 
      \begin{minipage}{8cm}
      \fontsize{8} {10pt}\selectfont
       {\bf Notes:} (a) Approximated values valid for the evolutionary status of 
       NGC 3293. (b) Known binaries and incompleteness effects were considered.
      \end{minipage}  
   \end{table} 

\subsubsection{Uncertainties in the LF and the IMF}

$\hspace{0.5cm}$
There are three main sources of uncertainties in the determination of both, the 
LF and the IMF: the field star correction, the incompleteness of the photometry 
and the unresolved binaries. Other effects of course do contribute, but the 
mentioned are the most difficult to assess. As for the removal of the star field 
contamination we are confident on the goodness of our method and the choice of 
the areas. Incompleteness is a bit more complex phenomenon to quantify. An 
analysis of the completeness factors was presented in section 2 and to remove 
in part this effect, we applied those factors to the amount of stars in each
bin of our LF. However, our conclusions may still be affected for incompleteness 
at the lowest luminosity bin. This uncertainty was minimised in our estimation 
of the LF that was set in the range from $-7$ to $+2$ and was also minimised in 
the computation of the IMF by excluding the lowest mass bin. Unresolved 
binaries, however, may influence the entire mass range. We will just describe 
the probable effect on our determination as the issue has been treated by many 
worker and detailed calculation have been already performed. Under the 
hypothesis that stellar masses of binary components are randomly distributed 
Sagar \& Richtler (1991) studied the mass range $2-14 \cal M_{\sun}$ finding 
that, if a fraction of $50 \%$ binaries is present in a given cluster, its IMF 
slope undergoes a flattening of $0.2$ for an initial slope of $1.5$. Lower 
initial slopes are less flattened according to these authors. Another analysis 
made by Kroupa et al. (1991, 1992) and Kroupa \& Gilmore (1992) also dealt with the 
point reaching the conclusion that if a Salpeter model is assumed, results lead 
to an apparent deficiency of low mass stars. This effect is probably the one 
producing the apparent deficiency of stars in the LF of NGC 3293 when compared 
to the overall LF of Phelps \& Janes (as marginally shown in Fig 9). \\
 
\section{The search for stars with $H\alpha$ emission} 

\begin{figure*}
   \centering
   \includegraphics[width=14cm]{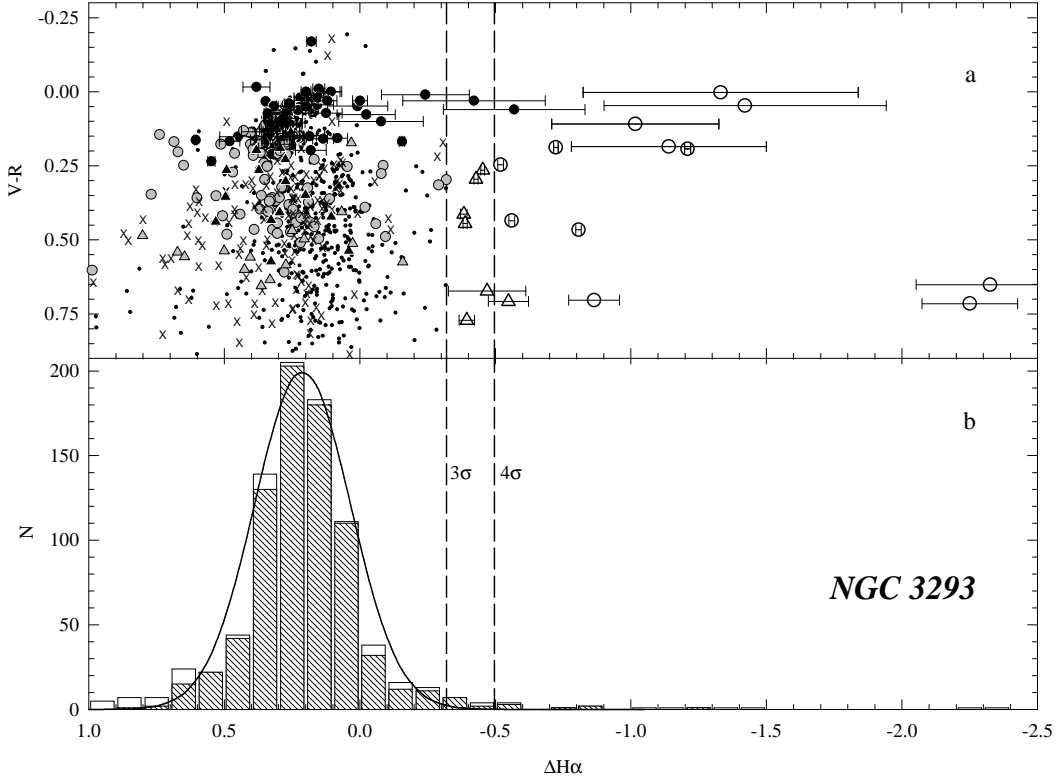}
      \caption{{} {\bf a)} $V-R$ vs. $\Delta~H_{\alpha}$ diagram. Symbols as in
      Fig 2a. Error bars from DAOPHOT task for `lm1' and $H_{\alpha}$ emission 
      stars are also indicated. {\bf b)} $\Delta~H_{\alpha}$ distribution. White
      histogram corresponds to all stars with $\Delta~H_{\alpha}$ index and
      the hatched one only includes those with index errors $< 0.1$. Solid
      curve is a Gaussian distribution function fitted to dashed histogram (see
      section 5). Dashed lines are $3 \sigma$ and $4 \sigma$ limits for
      $H_{\alpha}$ emission stars selection.}
      \label{Fig11}
\end{figure*}

   \begin{table*} 
      \centering
      \caption[]{{}Likely and probable $H_{\alpha}$ emission stars.} 
      \fontsize{8} {10pt}\selectfont 
      \begin{tabular}{rrrrrr@{.}lr@{.}lr@{.}lr@{.}lr@{.}lr@{.}l} 
      \hline 
          \multicolumn{17}{c}{{}Likely $H_{\alpha}$ emission stars.} \\
       \hline 
         $\#$ & $\#_{T}$ & $\#_{HM}$ & $X~~$ & $Y~~$ & \multicolumn{2}{c}{$V~~~$} & \multicolumn{2}{c}{$U-B~~$} & \multicolumn{2}{c}{$B-V~~$} & \multicolumn{2}{c}{$V-R~~$} & \multicolumn{2}{c}{$V-I~~$} & \multicolumn{2}{c}{{}$\Delta~H_{\alpha}~~~$} \\ 
      \hline 
         31 &  28 &  -  &  199.7 &  340.9 & 10&27   & -0&66    &  0&01    &  0&00    &  0&05    & -1&33 ::  \\
         63 &  -  & 277 &  196.9 &   85.5 & 12&02   & -0&28    &  0&22    &  0&18    &  0&39    & -1&14 ::  \\
         67 & 117 &  -  &  124.6 &  570.0 & 12&10   & -0&38 :  &  0&28    &  0&05    &  0&44    & -1&42 ::  \\
         85 &  -  & 264 &  208.8 &  170.1 & 12&34   & -0&48    &  0&10    &  0&11    &  0&23    & -1&02 ::  \\
        297 &  -  & 287 &  506.0 &  130.0 & 14&56   &  0&20    &  0&31    &  0&19    &  0&43    & -0&72     \\
        324 &  -  & 286 &  546.9 &  125.0 & 14&69   &  0&25    &  0&31    &  0&19    &  0&42    & -1&21     \\
        361 &  -  &  -  &  236.4 &  702.4 & 14&93   &   &      &   &      &  0&25    &   &      & -0&52     \\
        397 &  -  & 288 &  487.3 &  131.0 & 15&22   &  0&40    &  0&77    &  0&47    &  0&87    & -0&81     \\
        438 &  -  & 275 &  312.3 &   64.0 & 15&45   &  0&22    &  0&70    &  0&43    &  0&86    & -0&56     \\
        819 &  -  &  -  &  816.0 &  195.5 & 17&15   &   &      &   &      &  0&72    &   &      & -2&25 ::  \\
        989 &  -  &  -  &  602.4 &  765.1 & 17&72   &   &      &   &      &  0&65    &   &      & -2&33 ::  \\
       1544 &  -  &  -  &  771.5 &  519.2 & 19&62   &   &      &   &      &  0&70    &   &      & -0&86 :   \\
      \hline 
         \multicolumn{17}{c}{{}Probable $H_{\alpha}$ emission stars.} \\ 
      \hline 
         $\#$ & $\#_{T}$ & $\#_{HM}$ & $X~~$ & $Y~~$ & \multicolumn{2}{c}{$V~~~$} & \multicolumn{2}{c}{$U-B~~$} & \multicolumn{2}{c}{$B-V~~$} & \multicolumn{2}{c}{$V-R~~$} & \multicolumn{2}{c}{$V-I~~$} & \multicolumn{2}{c}{{}$\Delta~H_{\alpha}~~~$} \\       
      \hline 
        271 &  -  &  -  &  285.0 &  722.4 & 14&40   &   &      &   &      &  0&26    &   &      & -0&45     \\
        391 &  -  &  -  &  351.9 &  723.8 & 15&16   &   &      &   &      &  0&44    &   &      & -0&39     \\
        458 &  -  &  -  &  147.0 &  665.3 & 15&58   &  0&13 :  &  0&72 :  &  0&30    &  0&80 :  & -0&43     \\
        648 &  -  &  -  &  170.2 &  663.8 & 16&47   &  0&46 :  &  0&87 :  &  0&41    &  0&97 :  & -0&38     \\
        926 &  -  &  -  &  323.4 &   91.2 & 17&52   &  0&95 :: &  1&28    &  0&77    &  1&50    & -0&39     \\
       1448 &  -  &  -  &  667.3 &  722.6 & 19&20   &   &      &   &      &  0&71    &   &      & -0&55 :   \\
       1599 &  -  &  -  &  634.7 &  748.2 & 19&93 : &   &      &   &      &  0&67 :  &   &      & -0&47 ::  \\
      \hline 
      \end{tabular} 
      \begin{minipage}{16cm}
      $\hspace*{1.2cm}${\bf Notes:}
                       -$\#_{T}$ and $\#_{HM}$ means star numbers from TGHH80 
                        and HM82  respectively.\\
      $\hspace*{2.25cm}$-Colon (:) and double colon (::) indicate data with errors 
                        larger than  0.04 and 0.10 respectively.\\
      \end{minipage} 
   \end{table*} 
   
$\hspace{0.5cm}$
One of the aims of this work is the detection of stars with $H_{\alpha}$ 
emission. To assess whether a star shows $H_{\alpha}$ emission, we used the 
$H_{\alpha}(on)-H_{\alpha}(off)$ (hereafter $\Delta~H_{\alpha}$) index. Similar 
indices has been already used in other works (Adams et al. 1983; Sung et al. 
1998). We want to mention that due to poor weather conditions, the  seeing 
during the exposures of $H_{\alpha}$ frames was large ($2''-2''\!\!.5$) 
producing distortions in the final photometry of the  stars, especially those 
with close companions. That, probably, yielded the high  errors generated by 
DAOPHOT, especially in the $H_{\alpha}(off)$ frames. \\
 
Figure 11b shows the histogram (white) of $\Delta~H_{\alpha}$ values including 
all stars. The hatched one contains only stars with errors $< 0.10$ and was  
fitted with a Gaussian distribution function over the interval 
$-0.5 < \Delta H_{\alpha} < 0.9$ with mean value 
$\langle\Delta H_{\alpha}\rangle =  0.21$ and standard deviation 
$\sigma_{\Delta H_{\alpha}} = 0.18$. The mean value, which is assumed to 
correspond to non--emission objects, is close to the values found by Adams et 
al. (1983) in NGC 2264 ($0.30$) and NGC 7089 ($0.23$). However, the dispersion 
among our data is over twice the value $0.08$ found in NGC 2264 by them. 
Following Adams et al. reasoning, we identify a probable $H_{\alpha}$ emission 
star when its index value is $\Delta H_{\alpha} < -0.34$ (further than 
$3 \sigma$ from the mean value) and a likely emission star if it is 
$\Delta H_{\alpha} < -0.50$ (further than $4 \sigma$). All stars showing 
evidences of emission were plotted with white symbols in the photometric 
diagrams. It is interesting to mention the clear separation of stars with 
$H_{\alpha}$ emission when $\Delta H_{\alpha}$ values are plotted against the 
$V-R$ index (see Fig. 11a). We are aware that this procedure only detect stars 
with high $H_{\alpha}$ emission while most of PMS stars show weak $H_{\alpha}$ 
emission making their detection much harder (Preibisch \& Zinnecker 1999). \\
 
Nineteen stars, listed in Table 5, show evidences of having $H_{\alpha}$ 
emission. Their spatial locations in the NGC 3293 area (shown in Fig. 1) is 
interesting as eleven of them lie in the external border of the cluster and just 
another three are inside the cluster limits. From the spatial point of view 
there is no concentration of emission stars towards the cluster centre what 
suggests that the surrounding material of PMS stars in the cluster centre, that 
usually should produce $H_{\alpha}$ emission, has already been swept away by 
the powerful radiation field of the  most massive stars. Such a possibility has 
been suggested for  explaining the lack of emission stars in NGC 6231 (Sung et 
al. 1998) and in Upper Scorpius OB Association (Preibisch \& Zinnecker 1999). If 
this is true, the rest of PMS included in the cluster area should be weak--lined 
PMS stars only detectable as X--ray emitters (Montmerle 1996). \\
 
\section{Discussion} 

$\hspace{0.5cm}$
Uncertainties in the memberships and also in the stellar age assignment make 
difficult the interpretation of the star formation process in an open cluster. 
These uncertainties can lead to a wrong estimate of the star formation rate and 
therefore to a misinterpretation of the entire process. We attempted to reduce 
such uncertainties by verifying the reliability of the field star sample used to 
remove the contribution of field stars and proving that no mass segregation is 
present in NGC 3293. Moreover, when removing the field contribution from the 
CMDs of NGC 3293 we chose limits which define the PMS band in terms of both, 
extension and wide at a high level of credibility so that only a few stars must 
have been lost by this procedure. So, we do not expect our conclusions are 
seriously affected (qualitatively speaking). \\
 
As noticed above, the lack of coincidence between the mean stellar distribution 
of PMS stars and a given isochrone has been historically used as evidence 
against the coeval hypothesis (Iben \& Talbot 1966); in the present case there 
is a strong coincidence. There is a $2~Myr$ difference between the mean nuclear 
age and the mean contraction age, but it can be explained in terms of the 
various effects already mentioned. There is no chance either that the age 
difference between massive and PMS stars reaches the high value of $20~Myr$ 
indicated by HM82. That stars in NGC 3293 were all formed in a short period of 
time is additionally supported by the ``turn--on'' mass location at 
$\sim 2 \cal M_{\sun}$ (compatible with a contraction time ranging from $6$ to 
$12~Myr$) and by the most evolved star in the cluster, the red super--giant, 
which is $10 - 12~Myr$ old, thus equalising the mean age of PMS stars. \\
 
The computation of the SFR confirms the above suggestions in the sense that no 
total increasing SFR, as proposed in other clusters (e.g. NGC 6530 and NGC 2264, 
Iben \& Talbot 1966) is evident in NGC 3293. Indeed, the picture of this cluster 
could be similar to the re--interpretation of NGC 2264 and 6530 data (from Iben 
\& Talbot 1966)  made by Stahler (1985). This author found no obvious mass--age 
correlation but that the star formation in these two clusters took place in a 
period of time from $5$ to $12~Myr$. A similar results has been reported for the 
Upper Scorpius OB Association (Preibisch \& Zinnecker 1999; Preibisch et al. 
2002) and for the Scorpius--Centaurus OB Association (Mamajek et al. 2002) where 
was not found either evidences for a real age spread among PMS stars as the 
stellar formation proceeded during a short but intense burst of a few millions 
years, probably triggered by a supernova event. \\
 
\section{Conclusions}

$\hspace{0.5cm}$
We have investigated the open cluster NGC 3293 area with deep broad band and 
$H_{\alpha}$ photometry obtaining a picture of its main sequence structure down 
to $M_{V} \approx +4.5$. The picture demonstrates that the upper part of its 
sequence is composed by stars evolving off the ZAMS, the mid one mostly includes 
stars on the ZAMS yet and the lower main sequence consists of stars that are 
placed above the ZAMS becoming a PMS population. \\
 
Clear indications confirming the existence of a PMS population in this cluster 
are presented after a very careful removal of the field star contamination and 
also from the finding of $H_{\alpha}$ emission stars. Our analysis yielded that 
NGC 3293 is placed at a distance $d = 2750~pc$ and has a $4'\!.1$ angular radius 
($3.3~pc$). The absorption law affecting the cluster is normal although it is 
close to the northern part of the HII region NGC 3372. In this last particular 
place, abnormal extinction laws have been found (e.g. V\'{a}zquez et al. 1996; 
Tapia et al. 2002; Carraro et al. 2002). Following the common pattern of 
clusters in this galactic region, NGC 3293 shows, however, differential 
reddening surely produced by intracluster material. \\
 
Superposition of modern isochrones indicates this object has a nuclear age of 
$8~Myr$ and almost a similar mean value of $10~Myr$ was found among its 
faintest, PMS, stars. Therefore, the star formation in NGC 3293 appears to be 
coeval. Hints that star formation still continues (although at a probably low 
rate) in NGC 3293 is given by the detection of stars showing $H_{\alpha}$ 
emission. These stars are not concentrated towards the cluster centre but they 
tend to lie in the cluster periphery determined by star counts. The most likely 
explanation for this lack of concentration may be found in the fact that the 
radiation field of massive stars has swept away the material that surrounds the 
PMS stars. \\
 
The computation of the cluster LF did not show the strong dip in its structure 
nor the halo structure of faint stars, as previously suggested by HM82. As many 
other young clusters (Tr 14, V\'{a}zquez et al. 1996; NGC 6231, Baume et al. 
1999; or Pismis 20, V\'{a}zquez et al. 1995) we found that the IMF of NGC 3293 
has a slope value $x = 1.2 \pm 0.2$. This value is of the same order than the 
one found by Massey et al. (1995) who investigated many open clusters and 
associations in the Galaxy, the LMC and the SMC finding a mean slope values of 
$1.1\pm 0.1$. \\
 
{\it This article is partially based in the Digitized Sky Survey that was 
produced at the Space Telescope Science Institute under US government grant NAG 
W-2166. Original plate material is copyright the Royal Observatory Edinburgh and 
the Anglo--Australian Observatory. This research, also, has made use of the 
$SIMBAD$ database, operated at $CDS$, Strasbourg, France.}   
 
\begin{acknowledgements} 
{The authors acknowledge the financial support from the Facultad de Ciencias  
Astron\'{o}micas y Geof\'{\i}sicas de La UNLP and the IALP--CONICET. Special 
thanks are  given to Dr. Garrison for the allocation of telescope time at UTSO 
and to the CASLEO staff for the technical support. G.B. gratefully acknowledges 
the collaboration of the  Universit\`{a} di Padova (Italy) during a postdoctoral 
grant.} 
\end{acknowledgements}

\end{document}